\documentclass[letterpaper,twocolumn,english,superscriptaddress,aps,prb,floatfix,amsfonts,amssymb]{revtex4-1}
\usepackage[T1]{fontenc}
\usepackage[latin9]{inputenc}
\usepackage{textcomp}
\usepackage{amsmath}
\usepackage{amssymb}
\usepackage{graphicx}

\makeatletter

\@ifundefined{textcolor}{}
{%
 \definecolor{BLACK}{gray}{0}
 \definecolor{WHITE}{gray}{1}
 \definecolor{RED}{rgb}{1,0,0}
 \definecolor{GREEN}{rgb}{0,1,0}
 \definecolor{BLUE}{rgb}{0,0,1}
 \definecolor{CYAN}{cmyk}{1,0,0,0}
 \definecolor{MAGENTA}{cmyk}{0,1,0,0}
 \definecolor{YELLOW}{cmyk}{0,0,1,0}
}

\usepackage{epsfig}
\usepackage{float}
\usepackage{amscd}
\usepackage{bm}
\usepackage{psfrag}

\usepackage{babel}

\newcommand{\la}{\langle}
\newcommand{\ra}{\rangle}

\usepackage{babel}

\usepackage{babel}

\makeatother

\usepackage{babel}
\begin{document}
\title{Robust one-dimensionality at twin-grain-boundaries in MoSe$_{2}$}
\author{T. \v{C}ade\v{z}}
\thanks{These authors contributed equally to this work.}
\affiliation{Beijing Computational Science Research Center, Beijing 100193, China}
\affiliation{Center of Physics of University of Minho and University of Porto,
P-4169-007 Oporto, Portugal}
\author{L. Li}
\thanks{These authors contributed equally to this work.}
\affiliation{Department of Physics, National University of Singapore, Singapore
117551, Republic of Singapore}
\author{E. V. Castro}
\affiliation{Beijing Computational Science Research Center, Beijing 100193, China}
\affiliation{Center of Physics of University of Minho and University of Porto,
P-4169-007 Oporto, Portugal}
\affiliation{Departamento de Física e Astronomia, Faculdade de Ciências, Universidade
do Porto, 4169-007 Porto, Portugal}
\affiliation{CeFEMA, Instituto Superior Técnico, Universidade de Lisboa, Avenida
Rovisco Pais, 1049-001 Lisboa, Portugal}
\author{J. M. P. Carmelo}
\affiliation{Center of Physics of University of Minho and University of Porto,
P-4169-007 Oporto, Portugal}
\affiliation{Department of Physics, University of Minho, Campus Gualtar, P-4710-057
Braga, Portugal}
\affiliation{Boston University, Department of Physics, 590 Commonwealth Ave, Boston,
MA 02215, USA}
\affiliation{Massachusetts Institute of Technology, Department of Physics, Cambridge,
MA 02139, USA}

\date{24 December 2018}

\begin{abstract}
We show that 1D electron states confined at twin-grain-boundaries
in MoSe$_{2}$ can be modeled by a three-orbital tight binding model
including a minimum set of phenomenological hopping terms. The confined
states are robust to the details of the defect hopping model, which
agrees with their experimental ubiquity. Despite a valley Chern number
which is finite and opposite on both sides of the defect, there is
no topological protection of the confined states. This turns out to
be an essential feature to have only one confined electronic band,
in agreement with experiments, instead of two, as the bulk-edge correspondence
would imply. Modeling the confined state as a 1D interacting electronic
system allows us to unveil a mobile quantum impurity type behavior
at energy scales beyond the Tomonaga-Luttinger liquid with an interaction
range which extends up to the lattice spacing, in excellent agreement
with ARPES measurements. 
\end{abstract}
\maketitle

\section{Introduction}

One dimensional (1D) electronic systems are
the host of many interesting phenomena, including the possible condensed
matter realization of Majorana zero modes due to the non-trivial topology
of the electron states \citep{MZM2018}, the observation, due to electron
correlations \citep{Blumenstein-11}, of both low-energy Tomonaga-Luttinger
liquid (TLL) physics and higher-energy mobile quantum impurity model
(MQIM) behavior, beyond TLL \citep{Imambekov2012}, as well as the
observation of spin and charge separation at all energy scales \citep{MoSe-17},
to mention a few. In a three-dimensional world, one-dimensionality
is obviously not the rule. Fortunately, a variety of examples can
be found in nature (or synthesized) --- carbon nanotubes are a paradigmatic
example \citep{nanotubes2018}, but also semiconducting nanowires,
as for example InSb and InAs \citep{MZM2018,MZF+12}, and assembled
atom chains on surfaces \citep{Blumenstein-11,mjorFechain2014}, have
been on the spotlight recently, with prominent technological potential
in some cases.

\begin{figure}
\begin{centering}
\includegraphics[clip,width=0.98\columnwidth]{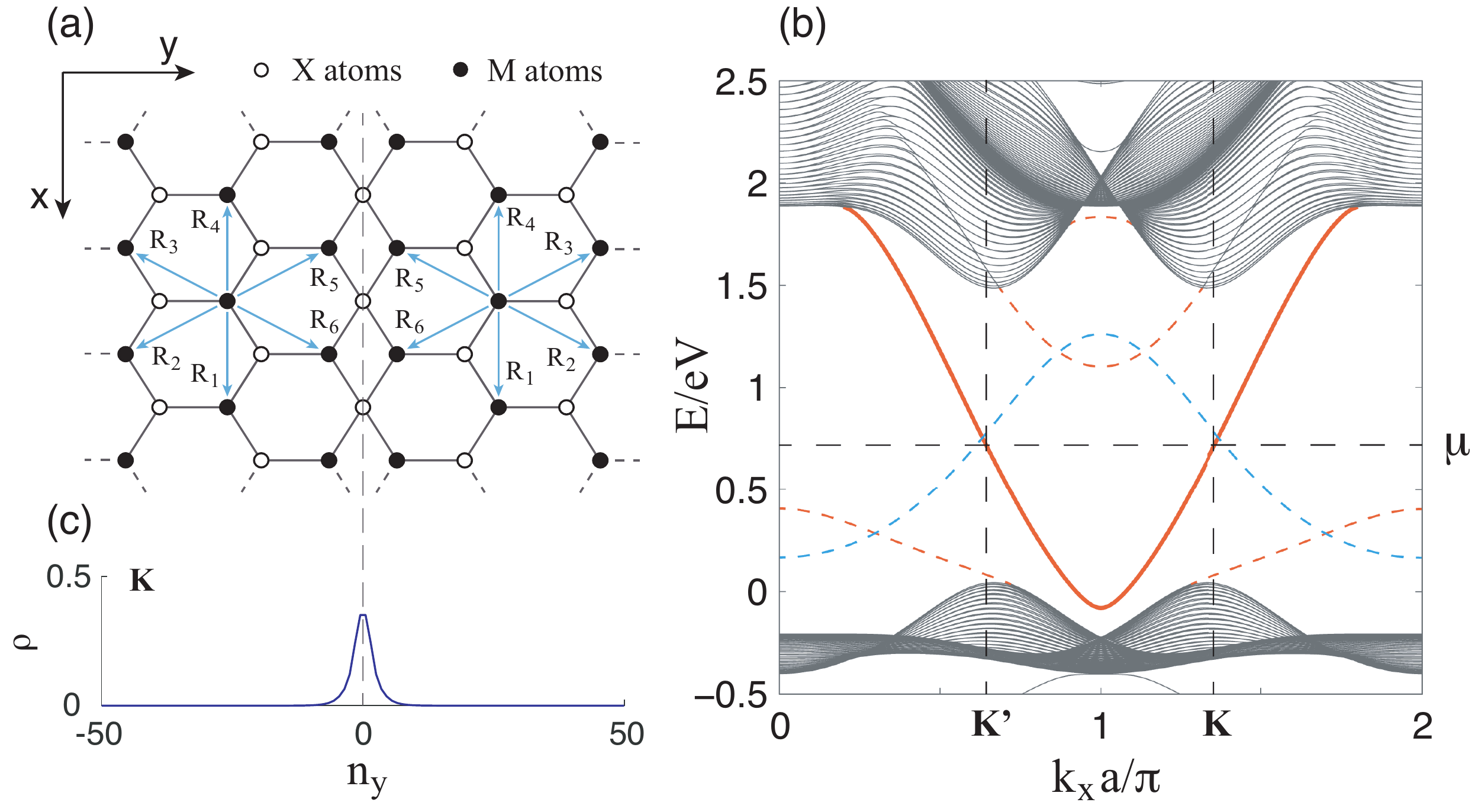} 
\par\end{centering}
\caption{\label{fig:latticeTB}(a) MX$_{2}$ lattice with a TGB defect along
the $x$ direction. (b)~Spectrum for a ribbon of MX$_{2}$ with a
TGB in the middle, obtained with a single hopping parameter to couple
the two sides of the TGB (see text). A band of electron states confined
at the TGB is shown as a thick (orange) line. Thin (black) lines are
bulk states and short-dashed (blue and orange) are other 1D states.
(c)~Probability density for an electron confined at the TGB in the
$K$-valley}
\end{figure}

The advent of two-dimensional materials \citep{NGPrmp}, in particular
the realization of a new class known as \emph{semiconducting transition
metal dichalcogenides} (TMDs) \citep{XXH14}, formula MX$_{2}$, where
M is a transition metal (ex. Mo, W) and X is a chalcogen (ex. S, Se)
\citep{wang2012electronics,yazievKis15}, allowed for a new type of
1D electron system: a confined state at the twin-grain-boundary (TGB)
defect shown in Fig.~\ref{fig:latticeTB}(a). The presence of such
1D states inside the bulk gap, in excess of 1~eV, has been clearly
demonstrated experimentally \citep{VanDerZande2013,Liu2014,Barja2016,MTBexp1}.
Their metallicity also became apparent, as well as intrinsic 1D behavior
such as a Peierls transition originating a charge density wave order
below $T\lesssim250\,\text{K}$, as well as spin and charge separation
characteristic of a correlated 1D system \citep{Barja2016,MoSe-17}.
In this paper, we show that the three-orbital tight binding (TB) model
of Ref.~\citep{xiao3bTB}, widely used to describe physics around
the gap edges in TMDs, can be used to describe the confined 1D states
at TGBs. A minimum set of phenomenological hoppings are included to
couple the two sides of the TGB. The induced in-gap states are robust
to the details of the defect hopping model, being present in its simplest
version where only nearest-neighbor (NN) hoppings between $d_{z^{2}}$
orbitals are allowed. The respective spectrum is showed in Fig.~\ref{fig:latticeTB}(b),
where a band of states localized at the TGB is clearly seen crossing
the gap. The localized nature of the states is depicted in Fig.~\ref{fig:latticeTB}(c),
where we show the probability density for a $K$-valley state. The
valley Chern number, which changes sign across the boundary and takes
values $C_{v}=\pm1$, does not warrant topological protection of the
1D states. This is crucial to stabilize a single band at the TGB,
in agreement with experiments and \emph{ab initio} simulations \citep{Zou2013,kisMTB2015,Barja2016},
as opposed to what would be implied by the Chern number change $|\Delta C_{v}|=2$ across
the TGB \citep{bernevigBook}. The stability of the single band is, however,
reminiscent of the Berry phase difference between the two sides of
the TGB \citep{Zhu2018}. 

Including interactions in the effective
1D system, and explicitly accounting for the effects of the finite
range of the interaction between the MQIM charge degrees of freedom,
improves the agreement with ARPES experiments beyond that reached
in Ref.~\citep{MoSe-17}.

The paper is organized as follows: In Sec.~\ref{TB}, we introduce
the tight-binding model used to describe the electronic properties
of the TGB. The continuum theory valid on both sides of the line defect
is discussed in Sec.~\ref{continuum}, where we also provide a detailed
topological analysis. The effect of the electron finite-range interactions within 
the line defects is studied in Sec.~\ref{ECLD}. In Sec.~\ref{conclusions}
the key results are summarized and some conclusions are drawn.
We also include two appendices: in Appendix~\ref{secap:continuum} we derive
the continuum theory; in Appendix~\ref{APA} 
some expressions useful for the discussion of the electron finite-range interactions
associated with  metallic states in the line defects are provided.

%%%%Tight-Binding

\section{Tight-binding analysis} \label{TB}

We model electrons in MoSe$_{2}$
using a M atom 3-orbital NN-TB Hamiltonian given by 
\begin{equation}
H_{0}=\sum_{i,\alpha}\sum_{\gamma,\gamma',\sigma}c_{i,\gamma,\sigma}^{\dagger}E_{\gamma,\gamma'}^{\sigma}(\mathbf{R}_{\alpha})c_{i+\mathbf{R}_{\alpha},\gamma',\sigma}\,,\label{eq:Hamiltonian_TB}
\end{equation}
where $c_{i,\gamma,\sigma}^{\dagger}$ is an electron creation operator
on lattice site $i$, M-atom orbital $\gamma=d_{z^{2}},d_{xy},\,d_{x^{2}-y^{2}}$,
spin $\sigma=\uparrow,\downarrow$, and $\mathbf{R}_{\alpha}$ with
$\alpha=1,\dots,6$ are the six vectors connecting NN atoms as shown
in Fig.~\ref{fig:latticeTB}(a). $E_{\gamma,\gamma'}^{\sigma}(\mathbf{R}_{\alpha})$
are hopping integrals as given in Ref.~\citep{xiao3bTB} for the
NN model \footnote{We use the hopping integrals obtained from the GGA DFT calculation
  of Ref.~\citep{xiao3bTB}}. We write the TB Hamiltonian, including the TGB, as
\begin{equation}
  H=H_{\mathrm{L}}+H_{\mathrm{R}}+H_{\mathrm{TGB}}\,,
  \label{eq:Hamiltonian_tot}
\end{equation}
with $H_{\mathrm{L}}\equiv H_{0}$ to the left of the TGB ($y<0$)
and $H_{\mathrm{R}}\equiv\sigma_{v}^{\dagger}H_{0}\sigma_{v}$ to
the right ($y>0$), where $\sigma_{v}$ is the reflection operator
associated to the mirror transformation $y\rightarrow-y$ {[}see Fig.~\ref{fig:latticeTB}(a){]},
and $H_{\mathrm{TGB}}$ couples left and right regions. $H_{\mathrm{R}}$
can be written as in Eq.~\eqref{eq:Hamiltonian_TB} with the NN hoppings
reversed {[}see Fig.~\ref{fig:latticeTB}(a){]},
so that the total Hamiltonian in Eq.~\eqref{eq:Hamiltonian_tot} respects
the apparent mirror symmetry of the system with respect to the line defect.
$H_{\mathrm{TGB}}$ is modeled in two ways: a simplified model, where only the NN hopping
between M-atom $d_{z^{2}}$ orbitals is allowed; and a more elaborated
model, where three NN hopping terms are allowed across the TGB.

\begin{figure}
\begin{centering}
\includegraphics[clip,width=0.98\columnwidth]{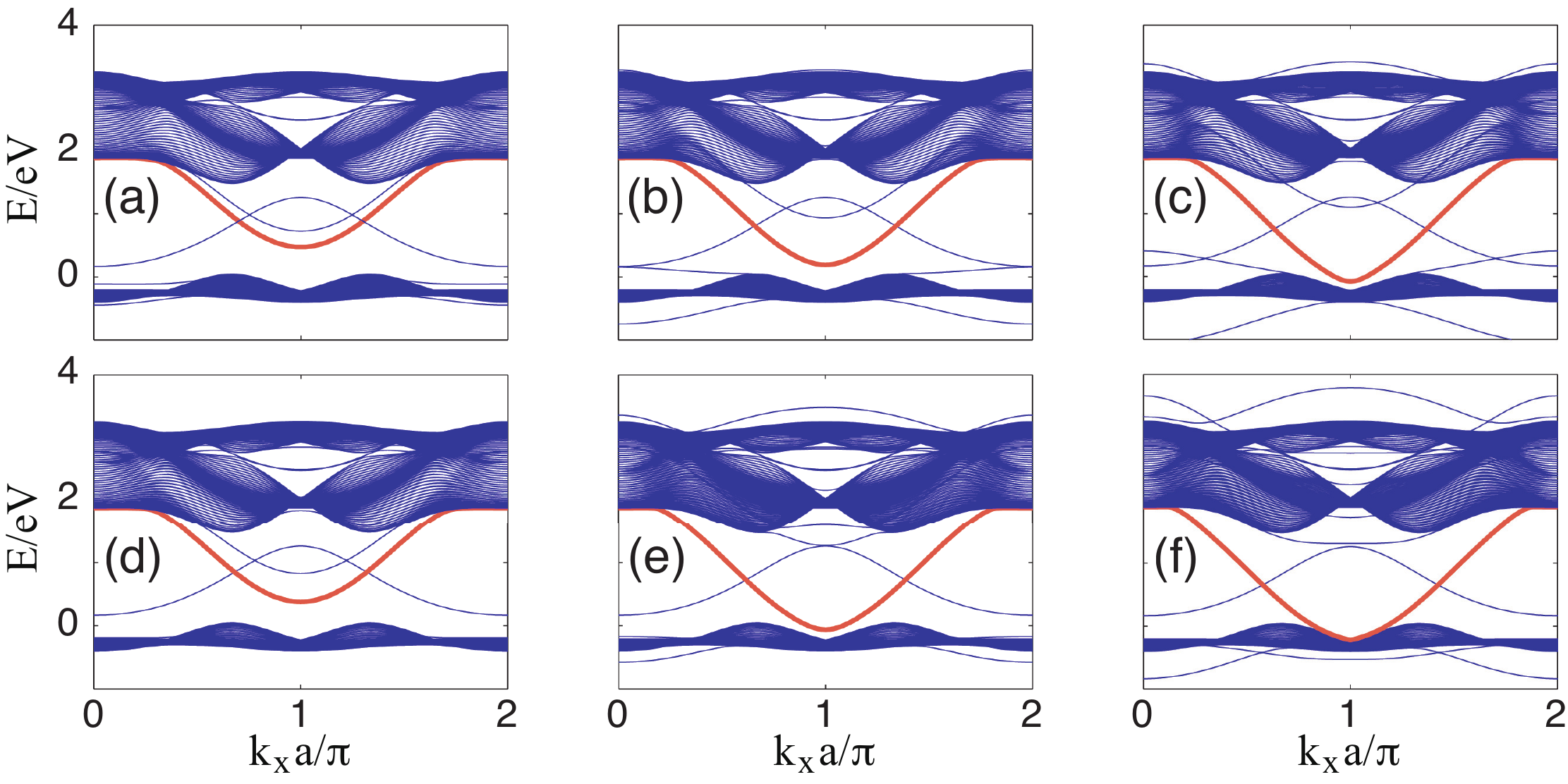} 
\par\end{centering}
\caption{\label{fig:TBcompar}Energy spectrum for a ribbon with a TGB in the
middle, obtained using the TB model with: (a)-(c)~only one NN hopping
between $d_{z^{2}}$ orbitals, $|\tilde{t}_{z^{2}}|=0.2,0.6,1.0\,\text{eV}$,
respectively; (d)-(e)~three NN hoppings involving $d_{z^{2}}$ and
$d_{x^{2}-y^{2}}$ (see main text), where $|\tilde{t}_{z^{2},x^{2}-y^{2}}|=0.2,0.6,1.0\,\text{eV}$,
respectively.}
\end{figure}

The results for the simplified model are shown in Fig.~\ref{fig:TBcompar}(a-c),
respectively for hopping values $|\tilde{t}_{z^{2}}|=0.2,0.6,1.0\,\text{eV}$,
where we considered a ribbon with translational invariance along $x$-direction
and $N_{y}=100$ unit cells in the $y$-direction, transverse to the
TGB. Panel~\ref{fig:TBcompar}(c) is the same as the one in Fig.~\ref{fig:latticeTB}(c).
In the latter, the dashed blue line corresponds to edge states localized
at the outer edges of the ribbon, so-called M-edges. These edge states,
present in all panels of Fig.~\ref{fig:TBcompar}, have been studied
elsewhere \citep{Bollinger2001,chen08,xiao3bTB} and will be ignored
here. In the limit $\tilde{t}_{z^{2}}=0$, the TGB is composed of
two uncoupled X-edges, which also support edge states \citep{chu2014Majorana,Li2016}.
In Fig.~\ref{fig:TBcompar}(a) it is seen that a finite $\tilde{t}_{z^{2}}$
lifts the degeneracy of the two X-edge states. On increasing $\tilde{t}_{z^{2}}$
{[}Figs.~\ref{fig:TBcompar}(b) and~\ref{fig:TBcompar}(c){]}, bonding
and anti-bonding states are formed. The bonding state is pushed down
in energy, particularly when the localization length is smaller ($k_{x}a\simeq\pi$),
and will be partially occupied. 

{\it Ab initio} calculations clearly show that the in-gap states localized
at the TGB are derived from M-atom orbitals \citep{Le2013, kisMTB2015}.
Within the 3-orbital NN-TB model adopted here, we have verified that including
hoppings involving the orbital $d_{xy}$ has little effect on the dispersion
of in-gap states, implying that the orbitals $d_{z^{2}}$ and $d_{x^{2}-y^{2}}$ are the most important
for the defect state. With this in mind, we developed a more realistic model for $H_{\mathrm{TGB}}$
considering three hoppings across the defect: direct hoppings $\tilde{t}_{z^{2}}$
and $\tilde{t}_{x^{2}-y^{2}}$, and a crossed term $\tilde{t}_{z^{2},x^{2}-y^{2}}$.
To reduce the number of free parameters, we fix the hopping ratios
to the values in the bulk, $\tilde{t}_{z{^{2}}}:\tilde{t}_{x^{2}-y^{2}}:-\tilde{t}_{z^{2},x^{2}-y^{2}}=t_{z\text{\texttwosuperior}}^{\mathrm{bulk}}:t_{x^{2}-y^{2}}^{\mathrm{bulk}}:t_{z^{2},x^{2}-y^{2}}^{\mathrm{bulk}}$.
The minus sign in $\tilde{t}_{z^{2},x^{2}-y^{2}}$ accounts for the
$\pi/2$ rotation of the hopping direction with respect to $\mathbf{R}_{1}$,
which is the reference for the hopping amplitudes in the bulk \citep{xiao3bTB}.
Figure~\ref{fig:TBcompar}(d-f) shows the spectrum for increasing
values of $|\tilde{t}_{z^{2},x^{2}-y^{2}}|=0.2,0.6,1.0\,\text{eV}$.
The results are very similar to those obtained with the single-hopping
model. Allowing for hoppings involving the $d_{xy}$-orbital does
not change significantly the results, which agrees with $d_{xy}$
minor role in TGB states.

In both models we allowed for hopping values $t\sim1\,\text{eV}$.
These are higher then bulk values \citep{xiao3bTB}, as a consequence of
the shorter NN distance between M-atoms on opposite sides of the TGB
($20\%$ smaller\citep{Le2013}).
We have deliberately ignored spin-orbit coupling (SOC) since TGB states
derive from the X-edge states, which are weakly affected by SOC. Intrinsic
spin-orbit coupling can be easily incorporated \citep{xiao2012,emmanueleTB13,asgariMoS2},
but only at very low temperatures will the spin-degeneracy assumption
break down.

%%%%Continuum and topology

\section{Continuum theory and topological considerations} \label{continuum}

\subsection{Low energy two-band model} \label{subsec:Low-energy-two-band}

A continuum theory describing the left ($y<0$) and right ($y>0$) regions {[}see
Fig.~\ref{fig:latticeTB}(a){]} can be derived from the three-orbital
TB model (see Appendix~\ref{secap:continuum}). The Hamiltonian reads 
\begin{equation}
\mathcal{H}_{\tau\mu}(\boldsymbol{q})=v\hbar(\tau q_{x}\sigma_{x}+\mu q_{y}\sigma_{y})+(\Delta+\beta q^{2})\sigma_{z}+\epsilon_{\mathrm{F}}\sigma_{0}\,,\label{eq:HlowE}
\end{equation}
where $\boldsymbol{q}=\boldsymbol{k}-\tau\boldsymbol{K}$ is the small
momentum with respect to valley $K$ ($\tau=+1$) or $K'$ ($\tau=-1$),
$\mu=+1$ on the left ($y<0$) and $\mu=-1$ on the right ($y>0$)
regions, and $\epsilon_{\mathrm{F}}$ is the chemical potential. The
Pauli matrices $\sigma_{i=x,y,z}$ act on the space of conduction
and valence band states at $\tau\boldsymbol{K}$, with $\sigma_{0}$
for the identity. For MoSe$_{2}$, the coefficients take the values:
$v\simeq5.6\times10^{5}\,\text{ms}{}^{-1}$, $2\Delta\simeq1.44\,\text{eV}$,
$\beta\simeq-3.01\,\text{eV}\,\text{Å}^{2}$, and $\epsilon_{\mathrm{F}}\simeq0.76\,\text{eV}$.
Apart from SOC, we are ignoring electron-hole asymmetry and trigonal
warping terms, which have much smaller coefficients (see Appendix~\ref{secap:continuum}).

Equation~\eqref{eq:HlowE} can be cast in the form
\begin{equation}
  \mathcal{H}_{\tau\mu}(\boldsymbol{q})=\boldsymbol{h}(\boldsymbol{q})\cdot\boldsymbol{\sigma}+
  \epsilon_{\mathrm{F}}\sigma_{0}\,,
  \label{eq:cont}
  \end{equation}
where $\boldsymbol{\sigma}$ is the vector of Pauli matrices. Equation~\eqref{eq:cont}
allows for straightforward topological analysis in terms of the valley Chern number, as done in the
following.

\subsection{Chern number}
Within the two-band continuum theory of the previous section, the
valley Chern number is defined by $C_{\tau,\mu}^{v}=\frac{1}{2\pi}\int_{-\infty}^{\infty}\int_{-\infty}^{\infty}\Omega_{\tau,\mu}^{v}(\boldsymbol{q})dq_{x}dq_{y}$,
with $\Omega_{\tau,\mu}^{v}$ the Berry curvature for the lower band
\citep{Xiao2010,Chiu2016}, 
\begin{equation}
  \Omega_{\tau,\mu}^{v}(\boldsymbol{q})=\frac{1}{2}\frac{\partial\boldsymbol{h}}{\partial q_{x}}\times\frac{\partial\boldsymbol{h}}{\partial q_{y}}.\frac{\boldsymbol{h}}{h^{3}}\,,\label{eq:Berry}
\end{equation}
with the vector $\boldsymbol{h}(\boldsymbol{q})$ as in Eq.~\eqref{eq:cont}.
After integration \citep{Lu2010}, we obtain 
\begin{equation}
C_{\tau,\mu}^{v}=\frac{1}{2}\tau\mu\left[\text{sign}(\Delta)-\text{sign}(\beta)\right].\label{eq:Chern}
\end{equation}
The dependence on $\Delta$ and $\beta$ is known \citep{Shen2012,rostamiTMDedge2016}:
for $\Delta>0$ and $\beta<0$, the case of TMDs, the system is topologically
non-trivial with $C_{\tau,\mu}^{v}=\tau\mu$. The dependence on the
valley index $\tau$ is required by time-reversal symmetry. The dependence
on $\mu$, which accounts for the position, left or right, with respect
to the TGB, is new and needs clarification.

For a Chern number change $\Delta C_{\tau}^{v}=|C_{\tau,+1}^{v}-C_{\tau,-1}^{v}|=2$,
we would expect two chiral modes per valley (per spin) running along
the boundary, as implied by the bulk-edge correspondence \citep{bernevigBook}.
These modes appear as bound states of the Hamiltonian $\mathcal{H}_{\tau\mu}(q_{x},y)$,
obtained from Eq.~\eqref{eq:HlowE} with $q_{y}\rightarrow-i\partial_{y}$
and $\mu\rightarrow\mu(y)$, where $\mu(y<0)=+1$ and $\mu(y>0)=-1$.
Close inspection shows that no bound state solution exists, contrary
to other 2D systems with domain walls \citep{MBM08,ZMM13,Vaezi2013}.
This is consistent with the absence of a gap closing associated with
a change of sign in $\mu$ (the spectrum $E_{\boldsymbol{q}}=\epsilon_{\mathrm{F}}\pm|\boldsymbol{h}(\boldsymbol{q})|$
is independent of $\mu$). The apparent discrepancy stems from the
transformation $y\rightarrow-y$ relating left and right regions,
which implies a reversal of the chirality of edge states, and thus
a sign change of the valley Chern number \citep{Araujo2014}, but
not a gap closing. The lack of topological protection, in contrast
to that found in topological band insulators \cite{Juricic2016},
is crucial to make our TB results compatible with experiments. Indeed,
for a filling $n=2/3$ (including spin), expected for the charge neutral
system, a single 1D band -- and not two -- crossing the Fermi level
was observed \citep{MoSe-17}. The stability of these states can be
linked to the 1D Berry phase difference between the two sides of the
TGB \citep{Zhu2018}, as discussed next.

\subsection{Low energy three-band theory} \label{subsec:lowE3b}

The inadequacy of the two-band theory of Sec.~\ref{subsec:Low-energy-two-band}
to describe bound states at the TGB can be understood within a continuum
three-band approximation. Such an approximation may be obtained by
writing the  momentum-space version of
$H_{0}$ defined in Eq.~\eqref{eq:Hamiltonian_TB}, and expanding
around the corners of the Brillouin zone (BZ). As is well known, 
right at the corner momenta $\boldsymbol{k}=\tau\boldsymbol{K}$ ($\tau=\pm1$),
we obtain Bloch states with well defined $z$-component orbital angular
momentum,\citep{xiao3bTB} 
\begin{align}
|\tau\boldsymbol{K},d_{0}\rangle= & |\tau\boldsymbol{K},d_{z^{2}}\rangle\nonumber \\
|\tau\boldsymbol{K},d_{+2\tau}\rangle= & \frac{1}{\sqrt{2}}\left[|\tau\boldsymbol{K},d_{x^{2}-y^{2}}\rangle+i\tau|\tau\boldsymbol{K},d_{xy}\rangle\right]\nonumber \\
|\tau\boldsymbol{K},d_{-2\tau}\rangle= & \frac{1}{\sqrt{2}}\left[|\tau\boldsymbol{K},d_{x^{2}-y^{2}}\rangle-i\tau|\tau\boldsymbol{K},d_{xy}\rangle\right]\,,\label{eq:Kstates}
\end{align}
with momentum states $|\boldsymbol{k},d_{\gamma}\rangle\equiv c_{\boldsymbol{k},\gamma}^{\dagger}|0\rangle$
dual to the $c_{i,\gamma}^{\dagger}|0\rangle$ states in Eq.~\eqref{eq:Hamiltonian_TB}.

\begin{figure}

\begin{centering}
\includegraphics[width=0.8\columnwidth]{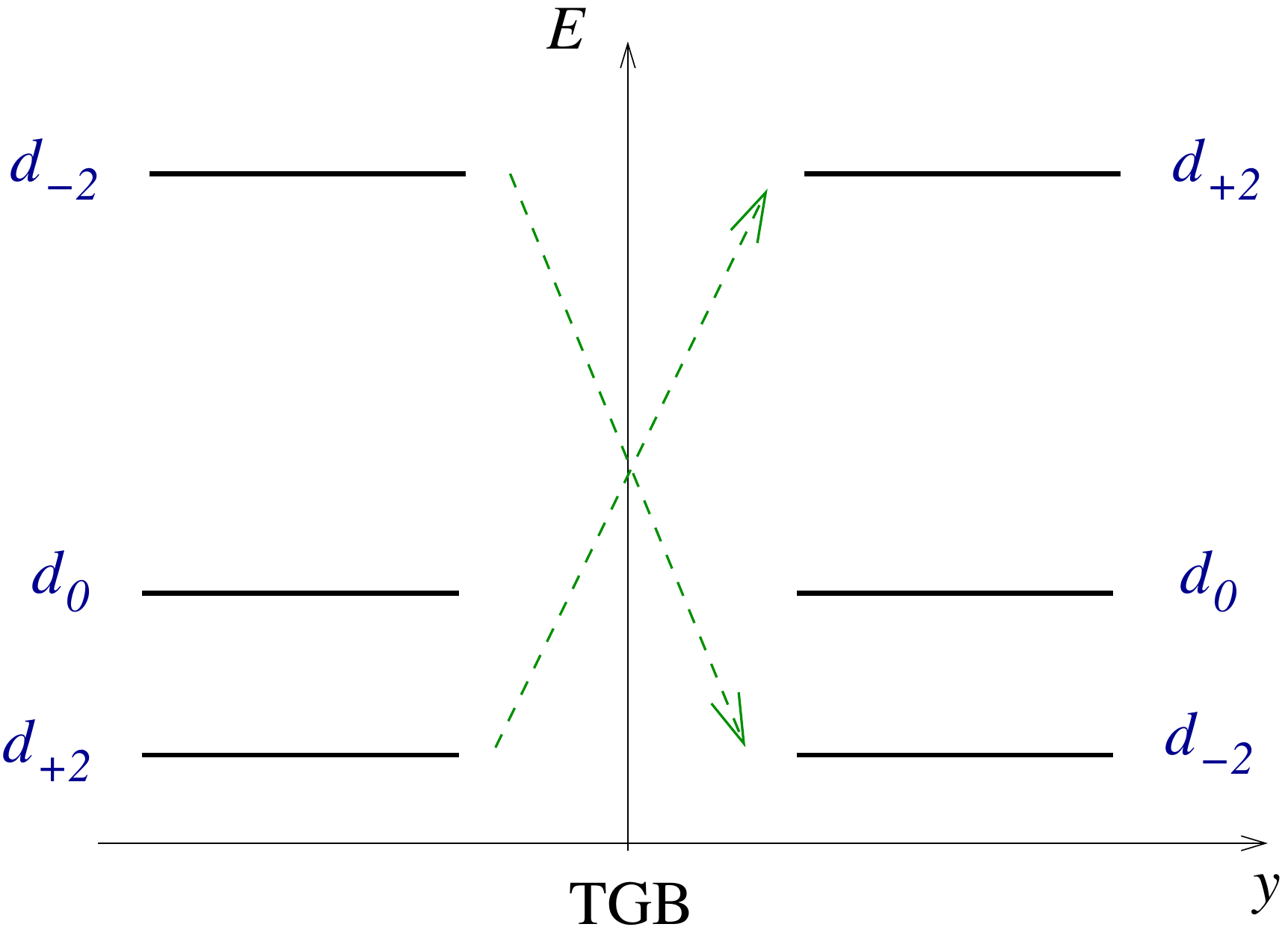}
\par\end{centering}
\caption{\label{fig:gapinv}Sketch of the band structure right at a single
valley momentum on both sides of the TGB. A band inversion involving
the highest and the lowest Bloch states $d_{\pm2}$ is apparent.}

\end{figure}

In the presence of the TGB a low energy three-band model can be invoked
far away from the line defect. On the $y<0$ side, Eq.~\eqref{eq:Hamiltonian_tot}
reduces to $H_{\mathrm{L}}$, while on the $y>0$ only $H_{\mathrm{R}}$
matters. Since $H_{\mathrm{L}}\equiv H_{0}$ and $H_{\mathrm{R}}$
is related to $H_{0}$ through a $y\rightarrow-y$ mirror transformation,
we can show that the three Bloch states given in Eq.~\eqref{eq:Kstates}
are eigenstates on both sides of the TGB right at the corner momenta
$\boldsymbol{k}=\tau\boldsymbol{K}$ (see Appendix~\ref{secap:continuum}
for details). However, the eigenergies are different, with a gap inversion
affecting the two states $d_{+2\tau}$ and $d_{-2\tau}$ {[}compare
Eqs.~\eqref{eq:3bLow1} and~\eqref{eq:3bLow2}{]}. Such gap inversion
is sketched in Fig.~\ref{fig:gapinv}.

The origin of the confined
1D states at the TGB may be traced back to the gap inversion involving
the valence and the highest bands described by the 3-orbital TB model
(see Fig.~\ref{fig:gapinv}).
A low-energy two-band approximation, where only the two lowest energy
states are considered, cannot capture this effect. This picture also
provides an understanding for why SOC effects are not important, as
these amount just to a small variation of the band energies, not affecting
the gap inversion.

\subsection{Berry phase} \label{secap:berry}

The state localized at the TGB is topologically originated at the
difference of Berry phase across the boundary\citep{Zhu2018}. 

The X- and M-edge states localize at the boundary along $x$, therefore
the system can be viewed as 1D lattice periodically modulated by a
parameter $k_{x}$, and the edge states can be described by a Berry
phase defined as 
\begin{eqnarray}
\gamma(k_{x})=i\oint dk_{y}\la u_{k_{x},k_{y}}|\partial_{k_{y}}|u_{k_{x},k_{y}}\ra=\pi,
\end{eqnarray}
with $u_{k_{x},k_{y}}$ the occupied state at 1/3 filling (without
spin), obtained by diagonalizing Eq.~\eqref{eq:Hk}\citep{Guo2015}.
A topologically nontrivial 1D insulating system is generally characterized
by a $\pi$ Berry phase, and has a pair of topologically protected
degenerate edge states localized at the two ends of the 1D chain.
However, the X- and M-edge states of TMDs have different energy dispersions
versus $k_{x}$, and $\gamma(k_{x})$ also varies continuously.

\begin{figure}
\begin{centering}
\includegraphics[clip,width=0.95\columnwidth]{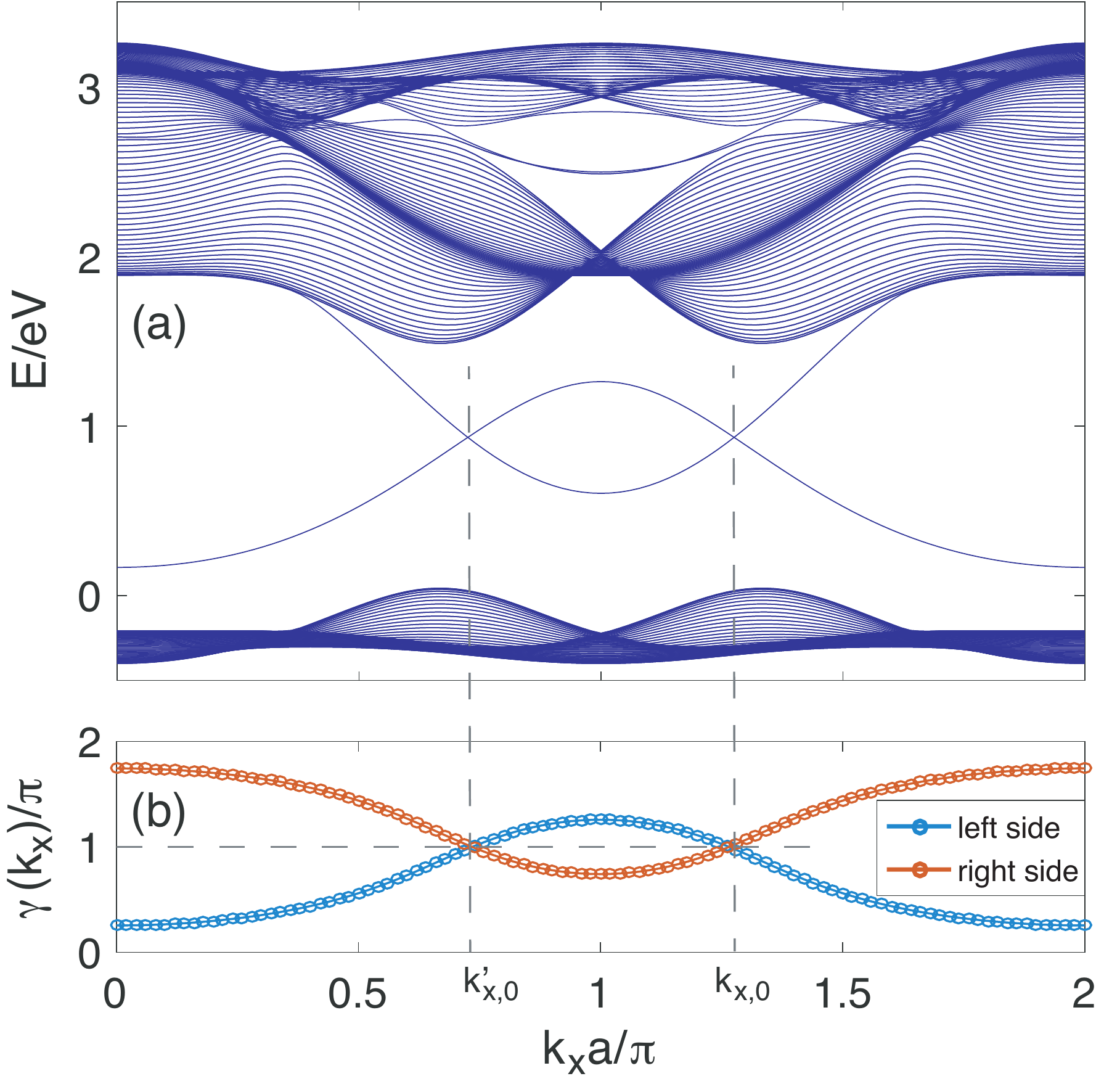} 
\par\end{centering}
\caption{\label{figS1} (a) The spectrum of the 3-band model of TMDs under
open boundary condition along $y$, and (b) the corresponding Berry
phase $\gamma(k_{x})$. The blue and red colors in (b) are for the
two halves with $y<0$ and $y>0$ respectively, in the system with
TGB defect as described in Fig.~\ref{fig:latticeTB}(a).}
\end{figure}

In Fig.~\ref{figS1} we illustrate the spectrum as a function of $k_{x}$
under open boundary condition along $y$, and the corresponding Berry
phase $\gamma(k_{x})$. The two edge states in Fig.~\ref{figS1}(a)
become degenerate at two certain values of $k_{x}=k_{x,0}$ (and $k'_{x,0}$),
corresponding to a Berry phase $\gamma(k_{x,0})=\pi$. The Berry phase
varies continuously away from $k_{x,0}$, thus the degeneracy of edge
states is lifted. Nevertheless, the edge states can be interpreted
as a continuation of the degenerate edge states at $k_{x,0}$, and
thus topologically originated at the $\pi$ Berry phase. In the system
with TGB that we consider in this work, the two sides with $y<0$ and
$y>0$ correspond to the Berry phase $\gamma(k_{x})$ and $-\gamma(k_{x})$
respectively, and the state localized at the TGB can be associated
with the difference of the Berry phase across the TGB \citep{Zhu2018}.

%%%%Correlations

\section{Effect of correlations within the line defects} \label{ECLD}

Consistent with the robust 1D nature of the metallic states in MoSe$_{2}$ line defects found here, 
the approach used by \textcite{MoSe-17} for a class of 1D correlated electronic lattice systems 
whose finite-range potential general properties are reported below
that applies to such states is a particular case of the general MQIM \citep{Imambekov2012}.
It uses a representation in terms of charge and spin particles that emerge in such systems
at all energy scales. The main effects of the electron repulsion between different sites are within that
approach in the interaction of the charge particles with the charge or spin hole mobile impurity created
under one-electron removal excitations. 

For the MoSe$_2$ line defects the effective range $R_{\rm eff}$ of the latter interaction
is small. Consistently, the studies of \textcite{MoSe-17} used $R_{\rm eff}= 0$. Here we account for the effects of 
higher-order charge-particle phase shift terms that contain $R_{\rm eff}$. We find that for the MoSe$_2$ 
line defects $R_{\rm eff}$ is of about one lattice spacing $a_0$. We confirm that using $R_{\rm eff}\approx a_0$ or 
$R_{\rm eff}\approx 0$ leads to theoretical predictions for such line defects ARPES peaks distribution within the 
experimental uncertainty. However, we find that accounting for the higher-order charge-particle phase shift 
terms and thus using $R_{\rm eff}\approx a_0$ improves the agreement with experiments. 

In this section we use generally units of lattice spacing $a_0$ one and Planck constant $\hbar$ one so 
that wave vectors are called momenta.

\subsection{The MQIM for finite-range interactions} \label{Transformation}

\begin{figure}
\centering{}\centerline{\includegraphics[width=0.60\columnwidth]{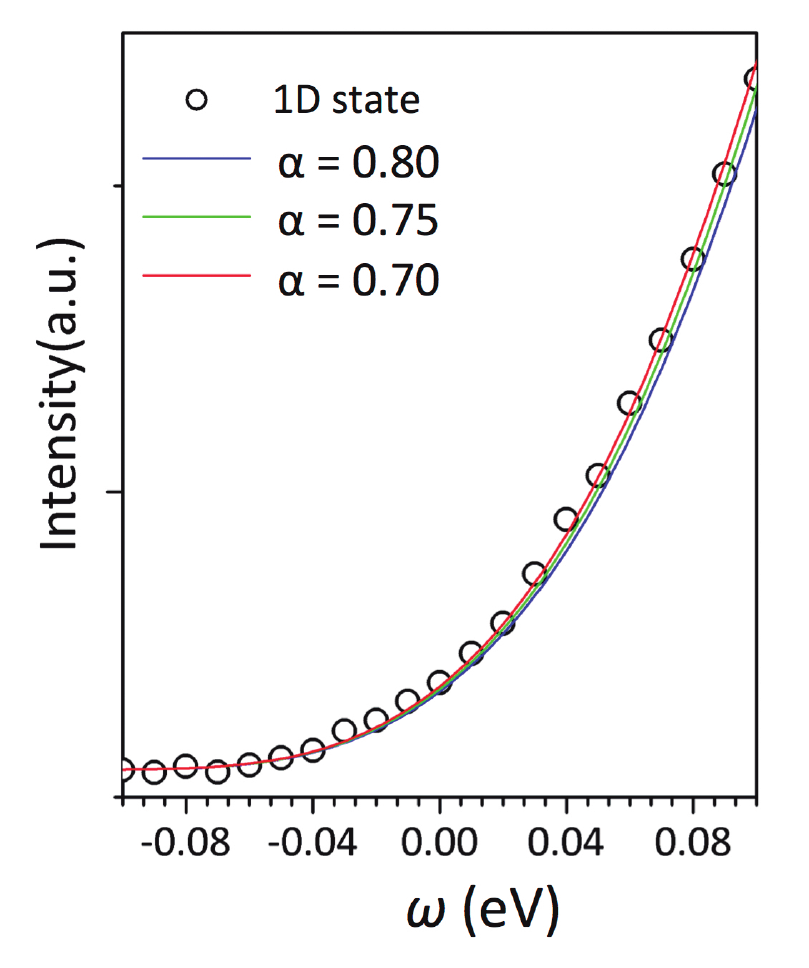}}
\caption{The suppression of the density of states of mirror twin grain boundaries
in monolayer MoSe$_{2}$ close to the Fermi-level measured at room
temperature and corresponding theoretical predicted power-law lines
for $\alpha=0.70$, $\alpha=0.75$, and $\alpha=0.80$. It is obtained
by plotting the angle integrated photoemission intensity as a function
of binding energy $\omega$. The experimental data are well fit for
$\alpha=0.75\pm0.5$ and thus with a corresponding uncertainty estimated
to be as large as $\pm0.05$. Source: Fig.~4(c) of \textcite{MoSe-17}.}
\label{figure3} 
\end{figure}

A decisive low-energy property of
1D metallic correlated systems is the low-energy power-law suppression
of the density of states (SDS) at the Fermi level. The experimental value of the corresponding power-law SDS exponent
$\alpha$ is typically equal to or larger than $1/2$ \citep{Blumenstein-11,Ohtsubo_15,MoSe-17}.
Figure \ref{figure3} displays the SDS of MoSe$_{2}$ line defects close to the Fermi-level
measured at room temperature (to avoid charge-density wave transition)
and corresponding analytical lines for SDS power-law exponent $\alpha=0.70$,
$\alpha=0.75$, and $\alpha=0.80$. 

It is known that the SDS exponent is such that $\alpha<1/8$ for the integrable 
correlated electronic models such as the
1D Hubbard model (1DHM) with onsite repulsion $U$ and transfer integral $t$
whereas an $\alpha>1/8$ stems from finite-range electron interactions 
in non-integrable models \citep{Schulz-90} whose range is at least of one lattice spacing.

According to the principle of emergence, the properties of a physical
system are mainly determined by how electrons are organized in it
\citep{Wen2017}. In the case of the correlated electronic systems
to which the MQIM applies \citep{Imambekov2012}, such an organization
gives rise to emerging fractionalized particles whose phase shifts
are imposed by mobile quantum impurities created under transitions
to excited states. 

The MQIM scheme used in the studies of Ref. \onlinecite{MoSe-17} accounted
for the leading-order term of an effective-range expansion of the charge-particle phase shift.
For the corresponding {\it leading-order} MQIM (MQIM-LO) \citep{MoSe-17}, the
emerging particles are the charge $c$ and spin $s$ (or $s1$) pseudofermions.
For simplicity, in this paper we call them charge $c$ and spin $s$ particles, respectively.
Both the general MQIM \citep{Imambekov2012} and the MQIM-LO 
used in the studies of \textcite{MoSe-17} provide 
accurate high-energy spectral function expressions beyond the low-energy Tomonaga-Luttinger liquid (TLL) 
theory \citep{Blumenstein-11}. For our purposes, by {\it high energy} we mean energy scales beyond the TLL limit.

Except for accounting for higher-order terms in the effective-range expansion of the charge-particle phase shift,
the expressions of the spectral-function quantities have for the {\it higher-order} MQIM (MQIM-HO) \citep{URT-18}
the same general form as for the MQIM-LO.
Within the MQIM-HO, the Hamiltonian that describes the 1D metallic states in the
corresponding class of electronic lattice systems is of the form,
\begin{eqnarray}
{\hat{H}} & = & -t\sum_{\sigma=\uparrow,\downarrow }\sum_{j=1}^{L}\left(c_{j,\sigma}^{\dag}\,
c_{j+1,\sigma} + c_{j+1,\sigma}^{\dag}\,c_{j,\sigma}\right) + \hat{V}_{R}
\nonumber \\
\hat{V}_{R} & = & \sum_{r=0}^{L/2-1}V_e (r)
\sum_{\sigma=\uparrow,\downarrow}\sum_{\sigma'=\uparrow,\downarrow}\sum_{j=1}^{L}\hat{\rho}_{j,\sigma}\hat{\rho}_{j+r,\sigma'} \, ,
\label{H}
\end{eqnarray}
where $\hat{\rho}_{j,\sigma} = \left(c_{j,\sigma}^{\dag}\,c_{j,\sigma} - {1\over 2}\right)$,
$V_e (0) = U/2$, $V_e (r) = U\,F_e (r)/r$ for $r>0$, and $F_ e(r)$ is a continuous decreasing 
screening function such that $F_e (0)\leq 1/4$, which at large $r$ vanishes as some inverse power of $r$,
$\lim_{r\rightarrow\infty}F_e (r)=0$. The microscopic interactions associated with the electronic potentials then
decay faster than $1/r$. Hence the Fourier transform of $V_e (r)$ does not diverge at $k\rightarrow 0$ and the
compressibility and sound velocity remain finite. 

The matrix elements in the one-electron spectral function involve phase shifts and the
charge parameter ${\tilde{\xi}}_c=\sqrt{2\tilde{K}_c}$ naturally related to them.
Its range for the present lattice systems is ${\tilde{\xi}}_c=\sqrt{2\tilde{K}_c}\in ]1/2,\xi_c]$.
Here $\tilde{K}_c$ is the usual TLL charge parameter and the
bare charge parameters $\xi_c \in ]1,\sqrt{2}[$ and $K_c$ refer to the 1DHM in which the 
model Hamiltonian, Eq. (\ref{H}), becomes in the ${\tilde{\xi}}_c\rightarrow\xi_c$ limit. 
For electronic density $n_e\in ]0,1[$ there is a $\xi_c\rightarrow {\tilde{\xi}}_c$ transformation 
\citep{MoSe-17} for each fixed value of $\xi_c$ and ${\tilde{\xi}}_c$ that maps the 
1DHM onto that model Hamiltonian, upon gently turning on $F_e (r)$. Consistent,
$\lim_{{\tilde{\xi}}_c\rightarrow\xi_c}F_e (r)\rightarrow 0$ for $r\in [0,\infty]$.
The MQIM-HO relies on that transformation. It transforms the 1DHM pseudofermion dynamical theory (PDT) \citep{Carmelo2018},
which for integrable models is equivalent to the MQIM \citep{Imambekov2012,Carmelo2018},
into the MQIM-HO that accounts for the electronic finite-range interactions of a class
of electronic lattice systems whose 1D metallic states are described by the 
model Hamiltonian, Eq. (\ref{H}).

As reported by \textcite{MoSe-17}, the $\xi_c\rightarrow {\tilde{\xi}}_c$ transformation 
gives rise to a continuous decreasing of the initial bare parameters $\xi_c = \sqrt{2K_c} \in ]1,\sqrt{2}[$ 
and $K_c = \xi_c^2/2 \in ]1/2,1[$. (Here $\xi_c = 1$ for $u=U/4t\rightarrow\infty$ and $\xi_c = \sqrt{2}$ for $u\rightarrow 0$, respectively.)
The resulting smaller renormalized parameter, ${\tilde{\xi}}_c=\sqrt{2{\tilde{K}}_{c}}$,
has values in the ranges ${\tilde{\xi}}_c =\sqrt{2{\tilde{K}}_{c}} \in ]1/2,1[$ and ${\tilde{\xi}}_c =\sqrt{2{\tilde{K}}_{c}} \in ]1,\xi_c]$.
The theory does not apply at the bare parameter $\xi_c= 1$ that refers to a non-metallic Mott-Hubbard insulating phase
at $n_e=1$ for $u>0$ and to $u\rightarrow\infty$ states whose spin configurations are all degenerated for $n_e\in ]0,1[$.
It also does not apply at ${\tilde{\xi}}_c=1$. Hence ${\tilde{K}}_{c} \in ]1/8,1/2[$ and ${\tilde{K}}_{c} \in ]1/2,K_c[$, so that, as 
expected \citep{Schulz-90}, ${\tilde{K}}_{c}>1/8$ for lattice correlated models. 

Importantly, upon decreasing ${\tilde{\xi}}_c$ from ${\tilde{\xi}}_c=\xi_c$ the initial 1DHM SDS exponent 
$\alpha_0=(2-\xi_c^2)^2/(8\xi_c^2)\in  ]0,1/8[$ continuously increases. Its expression is given by 
$\alpha = (2-{\tilde{\xi}}_c^2)^2/(8{\tilde{\xi}}_c^2)$. It has values in the corresponding intervals $\alpha\in [\alpha_0,1/8[$ 
and $\alpha\in ]1/8,49/32[$. The regime of more physical interest is ${\tilde{\xi}}_c \in ]1/2,1[$, for which $\alpha > 1/8$. 

For each chosen initial fixed 1DHM {\it finite} values $u = U/4t\in ]0,\infty[$ and $\xi_c = \xi_c (u,n_e)\in ]1,\sqrt{2}[$ 
where the electronic density varies in the interval $n_e\in ]0,1[$ there is {\it one} $\xi_c\rightarrow {\tilde{\xi}}_c$ 
transformation. Indeed, the system retains the
memory of $\xi_c$, and {\it both} $\xi_c$ and ${\tilde{\xi}}_c$ are MQIM-HO parameters that appear
in the expressions for physical quantities. The same applies to the scattering lengths
$a$ and ${\tilde{a}}$ considered below in Sec. \ref{SFPS}. The 1DHM initial interaction value $U$ remains under the
$\xi_c\rightarrow {\tilde{\xi}}_c$ transformation   
the interaction in both the onsite, $V_e (0) = U/2$, and $r>0$, $V_e (r) = U\,F_e (r)/r$,
parts of the electronic potencial in Eq. (\ref{H}).

\subsection{The one-electron removal spectral function and its exponents phase shifts} \label{SFPS}

\begin{figure}
\centering{}\centerline{\includegraphics[width=1\columnwidth]{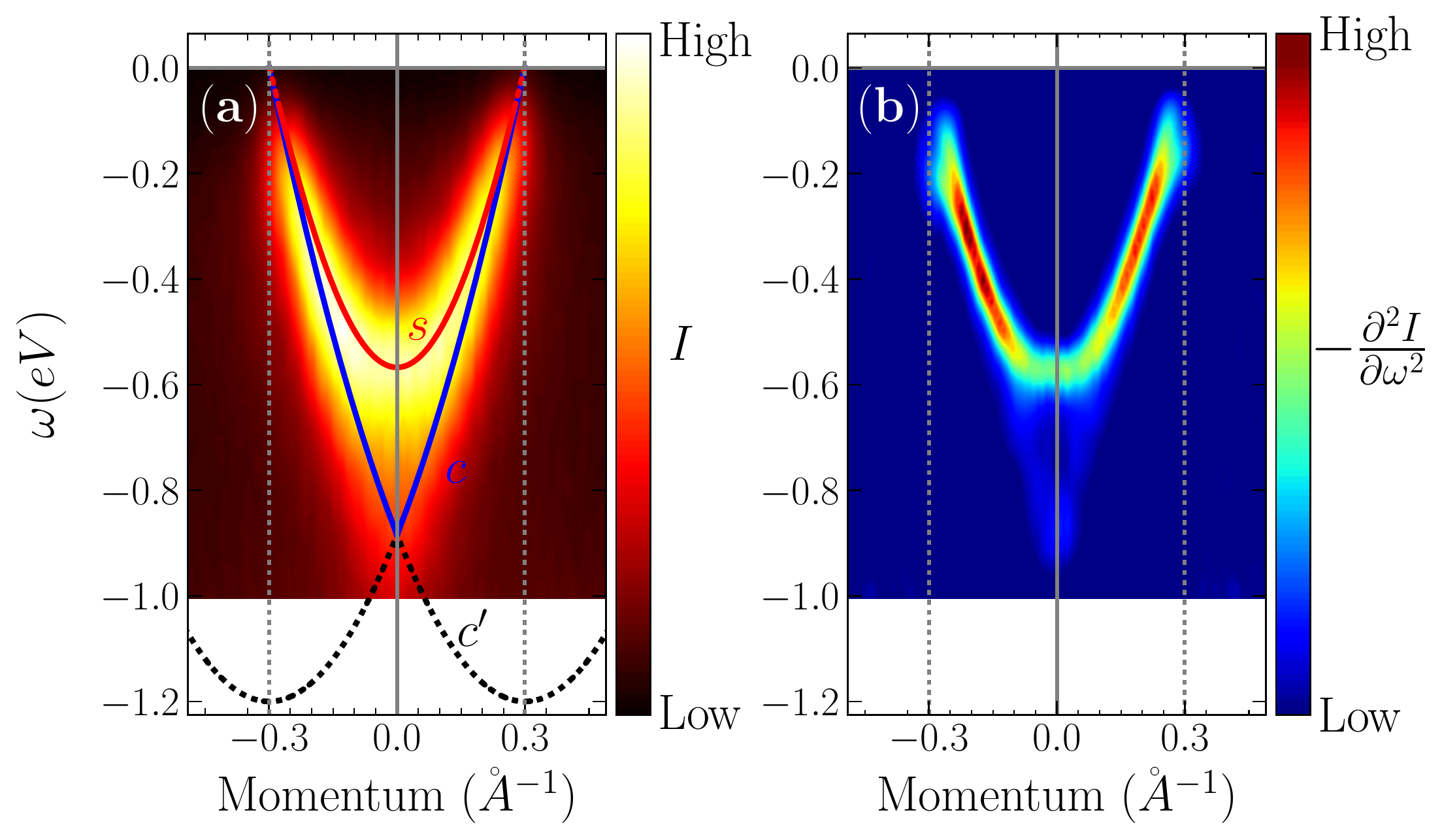}}
\caption{(a) Raw ARPES data image of MoSe$_{2}$ line defects with energy
versus momentum (k$_{//}$) along the $\overline{\Gamma_{01}}\,\overline{K}$
direction in the Brillouin zone plus the theoretical $c$, $c'$,
and $s$ branch-lines spectra \citep{MoSe-17} for $u=U/4t=0.18$,
transfer integral $t=0.58$ eV, and electronic density $n_{e}=2/3$.
The full and dashed lines refer to momentum ranges with negative and
positive exponents, respectively. (b) Second-derivative ARPES images.
Source: The experimental ARPES data are from \textcite{MoSe-17}.}
\label{figure4} 
\end{figure}

Within the MQIM-HO the one-electron removal spectral function in the $(k,\omega)$-plane
vicinity of three singular features called spin $s$ branch line and
charge $c$ and $c'$ branch lines, respectively, shown in Fig. \ref{figure4} (a)
has the form,
\begin{eqnarray}
& & {\tilde{B}} (k,\omega) = C_{s} ({\tilde{\omega}}_{s} (k)-\omega)^{{\tilde{\zeta}}_{s} (k)} 
\hspace{0.20cm}{\rm and}
\nonumber \\
& & {\tilde{B}} (k,\omega) \approx \sum_{\iota=\pm 1}(\iota)\,C_{\beta,\iota}
\nonumber \\
& \times &
{\rm Im}\left\{(-\iota)\left({\tilde{\omega}}_{\beta} (k)-\omega - {i\over 2\tau_{\beta} (k)}\right)^{{\tilde{\zeta}}_{\beta} (k)}\right\} \, ,
\label{Bkomega}
\end{eqnarray}
respectively, for small $({\tilde{\omega}}_{s}(k)-\omega)>0$ and 
$({\tilde{\omega}}_{\beta}(k)-\omega)>0$ where $\beta =c,c'$. Here $C_{s}$ and $C_{\beta,\iota}$
are $n_e$, $u=U/4t$, and ${\tilde{\xi}}_c$ dependent constants and $\omega<0$ are high energies. 

On the one hand, for ${\tilde{\xi}}_c\in [{\tilde{\xi}}_c^{\oslash},\xi_c]$ the 
$\beta =c,c'$ lifetimes $\tau_{\beta} (k)$ in Eq. (\ref{Bkomega}) is very large for the $k$ intervals for which 
the $\beta =c,c'$ exponents ${\tilde{\zeta}}_{\beta} (k)$ are negative, so that the expression given in that equation is nearly 
power-law like, ${\tilde{B}} (k,\omega) \propto ({\tilde{\omega}}_{\beta} (k)-\omega)^{{\tilde{\zeta}}_{\beta} (k)}$. 
The charge parameter value ${\tilde{\xi}}_c^{\oslash}=1/\xi_c$ is determined by that of the bare charge parameter $\xi_c$
and varies in the interval ${\tilde{\xi}}_c^{\oslash}\in [1/\sqrt{2},1[$.
Its smallest value ${\tilde{\xi}}_c^{\oslash}=1/\sqrt{2}$ refers to $\xi_c=\sqrt{2}$ and
$u\rightarrow 0$ whereas its non-reachable largest value ${\tilde{\xi}}_c^{\oslash}\rightarrow 1$ corresponds
to $\xi_c\rightarrow 1$ for $u\rightarrow\infty$. On the other hand, the effects of long-range interactions 
are stronger for ${\tilde{\xi}}_c\in ]1/2,{\tilde{\xi}}_c^{\oslash}]$.

The $\gamma=s,c,c'$ branch-line spectra ${\tilde{\omega}}_{\gamma}(k)$ in Eq. (\ref{Bkomega}) 
are provided in Eq. (\ref{equA1}) of Appendix \ref{APA}. They involve the $c$ and $s$ band energy dispersions given in Eq.
(\ref{equA2}) of that Appendix. The excitation momentum 
$k$ in those spectra argument are in Eq. (\ref{equA1}) of the same Appendix expressed in terms of the occupancies of the $c$ 
band momenta $q\in [-2k_F,2k_F]$ and $s$ band momenta $q'\in [-k_F,k_F]$ associated with the corresponding 
excited states. Here $2k_F = \pi\,n_e$.

\begin{figure}
\centering{}\includegraphics[width=0.98\columnwidth]{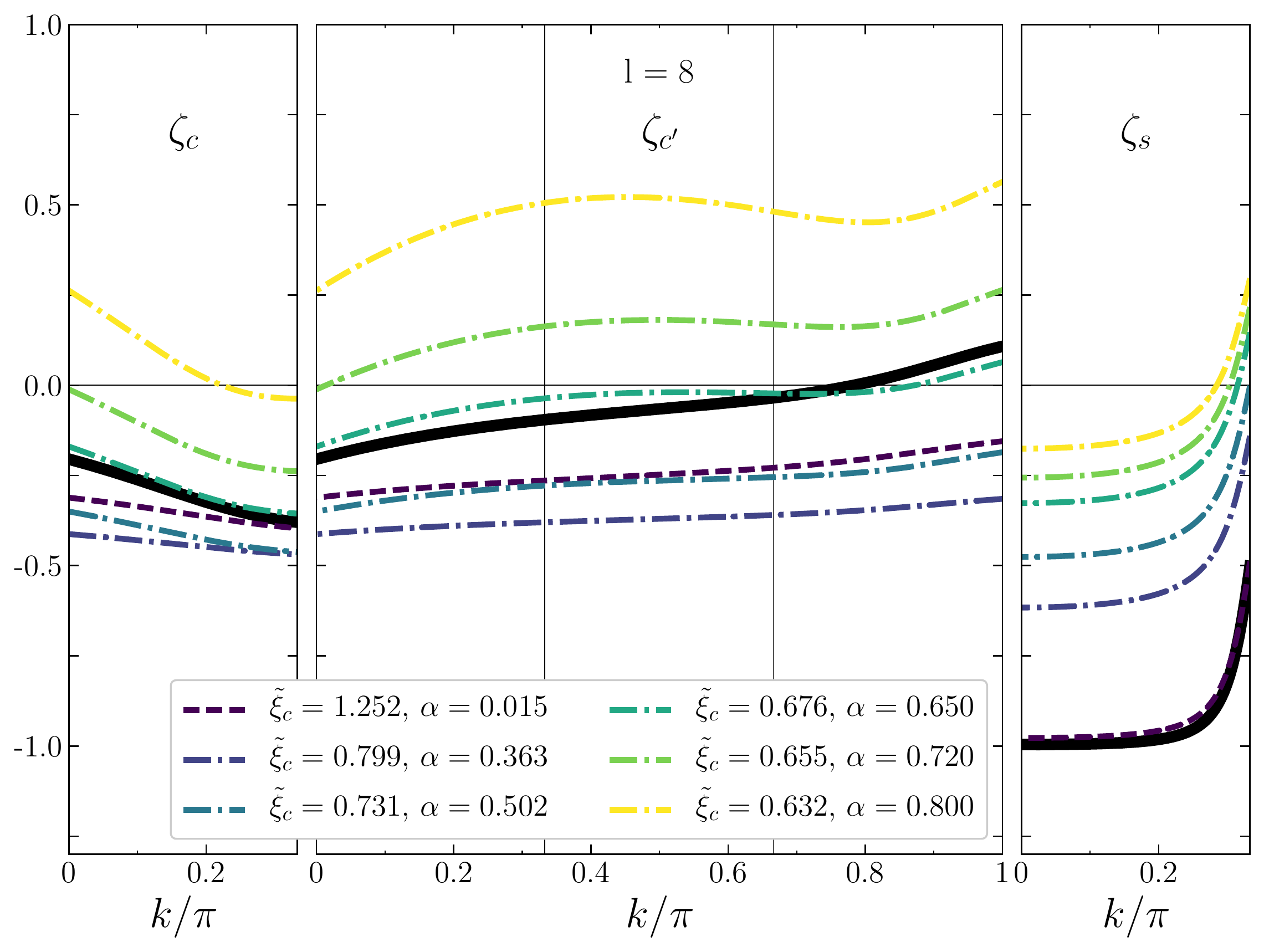} 
\caption{The exponents that control the line shape near the
MoSe$_{2}$ line defects ARPES peaks and corresponding theoretical
$c$, $c'$, and $s$ branch lines, respectively, in Fig. \ref{figure4} (a). They
are here plotted as a function of the momentum $k$ for $u=0.18$,
$n_{e}=2/3$, $l=8$, and different ${\tilde{\xi}}_{c}$ and thus
$\alpha$ values. The black solid lines refer to the conventional
1D Hubbard model ($\alpha_{0}=0.0011336$ and $\xi_{c}=1.367$) and
the red dashed and the blue (dashed-dotted and full) lines to $\alpha<1/8$
and $\alpha>1/8$ values, respectively. The $c$ line, $c'$ line, and $s$ line whose
negative exponents ranges agree with the ARPES $(k,\omega)$-plane
peaks in  Fig. \ref{figure4} (a) are those whose $c'$ branch-line exponent crosses
zero at $k/\pi=0$. For such lines, ${\tilde{\xi}}_c=0.655$, $\alpha=0.72$ and $R_{\rm eff}=1.01$
in units of lattice spacing. The ${\tilde{\xi}}_{c}$ value below which
the effects of long-range interactions become stronger is ${\tilde{\xi}}_c^{\oslash}=1/\xi_c=0.731$.}
\label{figure5}
\end{figure}

Moreover, $\tau_{c} (k)$ and $\tau_{c'} (k)$ are in Eq. (\ref{Bkomega}) 
large charge hole mobile impurity lifetimes mentioned above. They are
associated with the relaxation processes discussed below and
the expressions of the $\gamma=s,c,c'$ exponents ${\tilde{\zeta}}_{\gamma}(k)$ in 
Eq. (\ref{Bkomega}) are given in Eq. (\ref{equA3}) of Appendix \ref{APA}. They involve the charge parameter 
${\tilde{\xi}}_c$ and the $c$ particle phase shifts ${\tilde{\Phi}}_{c,s}(\iota 2k_F,q')$ and
${\tilde{\Phi}}_{c,c}(\iota 2k_F,q)$ where $\iota =\pm 1$. They are the phase shifts in units of $2\pi$ imposed on
a $c$ particle of momentum $\iota 2k_F=\pm 2k_F$ by a $s$ (spin) and $c$ (charge) hole mobile impurity
created at momentum $q'$ and $q$, respectively, under one-electron removal excitations.
Such exponents expressions also involve phase shifts ${\tilde{\Phi}}_{s,s}(\pm k_F,q')$ and ${\tilde{\Phi}}_{s,c}(\pm k_F,q)$ imposed 
on the $s$ particles by a $s$ (spin) and $c$ (charge) hole
mobile impurity, respectively. They remain hidden because they are invariant under the 
$\xi_c\rightarrow {\tilde{\xi}}_c$ transformation and due to the $SU(2)$ symmetry are interaction, density, and 
momentum independent, as given in Eq. (\ref{equA4}) of Appendix \ref{APA}.
The exponents ${\tilde{\zeta}}_{\gamma}(k)$ are plotted in Fig. \ref{figure5} as a function
of the excitation momentum $k$ for $u=U/4t=0.18$ and electronic density $n_{e}=2/3$.

In the low-energy TLL regime and in the cross over regime to it that refer to small-energy 
regions near the $(k,\omega)$-plane points $(\pm k_F,\omega)$ for the $s$ and $c$ branch 
lines and $(\pm 3k_F,\omega)$ for the $c'$ branch line, the corresponding exponents expressions 
are different from those provided in Eq. (\ref{equA3}) of Appendix \ref{APA}. Fortunately,
the ARPES peaks studied here refer to higher energy scales at which the latter exponents
apply.

The microscopic processes that control the weight distribution near 
the $\gamma=s,c,c'$ branch line singularities of the one-electron removal spectral function at 
$k$ domains for which the exponents ${\tilde{\zeta}}_{\gamma}(k)$ in Eq. (\ref{equA3}) of Appendix \ref{APA}
are negative refer to creation of one hole in the $c$ band and one hole in the $s$ band. 
Specifically, in the case of the $s$ branch line the $s$ band hole is created away from the
corresponding Fermi points $\pm k_F$ whereas the $c$ band hole is created at one of
that band Fermi points $\pm 2k_F$. The charge $c$ and $c'$ branch lines result from processes 
under which the $c$ band hole is created away from the corresponding Fermi points $\pm 2k_F$
and the $s$ band hole is created at one of its bands Fermi points $\pm k_F$. Furthermore, the $c$ band
discrete momenta are all shifted by $\pi/L$ or $-\pi/L$ whereas those of the $s$ band 
are not. This leads to an overall macroscopic shift of momentum $2k_F$ or $-2k_F$, respectively, 
which originates from the shifting of the whole $c$ band occupied sea. 

Such a shifting is behind the existence of two independent charge branch lines. The parts of these two branch lines 
that connect the point $(k,\omega)=(-k_F,0)$ in Fig. \ref{figure4} (a) to a $k=0$ 
finite-$\omega$ point and the latter point to $(k,\omega)=(k_F,0)$ are here and in the figure
called the {\it $c$ branch line}. The remaining parts of the charge branch lines that connect
the point at $k=0$ and finite $\omega$ to the $(k,\omega)=(-3k_F,0)$ and $(k,\omega)=(3k_F,0)$ 
points, respectively, are called the {\it $c'$ branch line}. (Because one finds below that for the
parameters suitable to the theoretical description of the ARPES in the MoSe$_2$ line defects
there are no singularities in the $c'$ branch line, in Fig. \ref{figure4} (a) only part of its $k$ range
is included.)

{\it Only} the charge hole {\it or} spin hole, respectively, that is created away from the
corresponding Fermi points is called a {\it mobile impurity}. The high-energy MQIM-HO charge hole quantum mobile impurity 
and spin hole quantum mobile impurity become in the low-energy limit the usual TLL holon and spinon, respectively.

On the one hand, since the $c$ and $c'$ branch lines lie in the spectral-weight continuum,
in their vicinity the spectral-function expression given in Eq. (\ref{Bkomega}) is
for the charge parameter range ${\tilde{\xi}}\in ]1/2, {\tilde{\xi}}_c^{\oslash}]$ for which
the effects of the finite-range interactions are stronger such that their power-law 
singularities are slightly broadened by weak charge hole mobile impurity
relaxation effects associated with large lifetimes $\tau_{c} (k)$ and $\tau_{c'} (k)$.
However, they remain sharp peaks for the $k$ ranges for which the exponents
${\tilde{\zeta}}_{c}(k)$ and ${\tilde{\zeta}}_{c'}(k)$, respectively, given in Eq. (\ref{equA3}) of Appendix \ref{APA}
are negative. For ${\tilde{\xi}}> {\tilde{\xi}}_c^{\oslash}=1/\xi_c$ the relaxation effects are
much weaker and the above reported $\beta =c,c'$ branch lines singularities power-law behavior 
${\tilde{B}} (k,\omega) \propto ({\tilde{\omega}}_{\beta} (k)-\omega)^{{\tilde{\zeta}}_{\beta} (k)}$
is a good approximation for their expression given in Eq. (\ref{Bkomega}). What matters for the description of the MoSe$_2$
line defects ARPES peaks distribution reported below in Sec. \ref{ARPES} is not though the precise form 
of the theoretical spectral function near its peaks but rather the $k$ ranges for which its
exponents are negative. They provide precise and valuable information on the predicted
location of such peaks in the $(k,\omega)$ plane. 

On the other hand, the $s$ branch line coincides with an edge of
support of the spectral function that limits the finite-weight region.
Then the scattering processes allowed by energy and momentum conservation 
ensure that the expression of the exponent ${\tilde{\zeta}}_{s} (k)$ in Eq. (\ref{Bkomega}) is exact.

The $s$ particle energy dispersion remains invariant under the $\xi_c\rightarrow {\tilde{\xi}}_c$ transformation. 
The $c$ particle energy dispersion bandwidth of the 
occupied sea increases slightly \citep{URT-18}. (See Eq. (\ref{equA2}) of Appendix \ref{APA} where 
${\tilde{\varepsilon}}_c (q)$ and ${\tilde{\varepsilon}}_s (q')=\varepsilon_s (q')$
are the MQIM-HO energy dispersions and $\varepsilon_c (q)$ and $\varepsilon_s (q')$ those 
associated with the bare limit, ${\tilde{\xi}}_c =\xi_c$, that refers to the 1DHM.)
That the spin spectra remain invariant under finite-range interactions whereas
the charge spectra bandwidth and charge Fermi velocity are increased upon
increasing the interactions range, is also known from numerical studies \citep{Hohenadler-12}.
(See charge and spin spectra in Fig. 7 of that paper and corresponding discussion.)

However, the major effects of the finite-range interactions are 
on the one-electron matrix elements between the ground state and the excited states. 
In the representation in terms of charge and spin particles such effects lead to a renormalization 
of the phase shifts of the charge particles imposed by the charge and spin hole mobile 
impurities created under transitions to the one-electron removal excited states \citep{MoSe-17}.
The renormalization of the phase shifts $2\pi{\tilde{\Phi}}_{c,s}(\pm2k_{F},q')$ and $2\pi{\tilde{\Phi}}_{c,c}(\pm 2k_{F},q)$ 
appearing in the exponents expressions, Eq. (\ref{equA3}) of Appendix \ref{APA},
under the $\xi_c\rightarrow {\tilde{\xi}}_c$ transformation leads to Eq. (\ref{PhaseShifts})
of Appendix \ref{APA}.

The MQIM-HO phase shift term $2\pi{\tilde{\Phi}}_{c,c}^{R_{\rm eff}} (k_r)$ in that equation is absent from
the 1DHM as it emerges from finite-range interactions higher-order effects beyond the renormalization 
factor $[\xi_c({\tilde{\xi}}_c -1)^2]/[{\tilde{\xi}}_c(\xi_c -1)^2]$
of the phase-shift term $2\pi{\tilde{\Phi}}_{c,c}^{{\tilde{a}}} (\pm 2k_F,q)$. 
(That term has not been considered in the MQIM-LO of \textcite{MoSe-17}.)

Such higher-order effects result from the potential $V_{c}(x)$ associated 
with the interaction of the charge $c$ particle and the charge hole mobile impurity at spatial distance $x$, 
which is induced by the electronic potential $V_e (r)$ in Eq. (\ref{H}). For the class of MQIM-HO electronic potentials,
that induced potential $V_{c}(x)$ vanishes for large $x$ as $V_{c}(x)=-C_{c}/x^{l}$. Here $l\geq6$ is an integer
determined by the large-$r$ behavior of $V_{e}(r)$, $C_{c}=(2r_{l})^{l-2}/\mu$,
$r_{l}$ is a length scale (van der Waals length for $l=6$), and
$\mu$ is the reduced mass \citep{URT-18}.

The phase-shift term $2\pi{\tilde{\Phi}}_{c,c}^{R_{\rm eff}} (k_r)$ expression \citep{URT-18} involves the effective range $R_{\rm eff}$
of the interactions between the $c$ particles at and near the $c$ band Fermi points $\pm 2k_F$
and the charge hole mobile impurity created under one-electron removal excitations at $c$ band momenta
$q$ away from the $c$ band Fermi points. It is a function of the corresponding relative momentum,
$k_r = q\mp 2k_F$, such that $\vert k_r\vert \in [k_{Fc}^0,4k_F[$. The use of standard scattering 
theory for potentials with large-$x$ behavior $V_{c}(x)=-C_{c}/x^{l}$ where $l\geq6$
leads to a $R_{\rm eff}$ effective range expression that involves the ratio ${\tilde{a}}/a$
of the scattering length ${\tilde{a}}$ corresponding to the renormalized charge parameter
${\tilde{\xi}}_c$ value and the bare scattering length $a$ associated with the $n_e$ and
$u=U/4t$ dependent bare charge parameter $\xi_c$ value, respectively \citep{URT-18}. 

\subsection{Application to the ARPES peaks distribution} \label{ARPES}

The higher-order charge-charge interaction effects associated with 
the phase-shift term $2\pi{\tilde{\Phi}}_{c,c}^{R_{\rm eff}} (k_r)$
play an important role in the one-electron spectral properties of 1D metallic states 
as those in a bismuth-induced anisotropic structure on 
indium antimonide [Bi/InSb(001)] whose effective range $R_{\rm eff}$ 
can reach values $R_{\rm eff}\approx 17$ in units of lattice spacing \citep{URT-18}. 

The studies of \textcite{MoSe-17} on the MoSe$_{2}$ line defects considered that $R_{\rm eff}= 0$
and thus that $2\pi{\tilde{\Phi}}_{c,c}^{R_{\rm eff}} (k_r)=0$
in the expression of the phase shift $2\pi{\tilde{\Phi}}_{c,c} (\pm 2k_F,q)$
in Eq. (\ref{PhaseShifts}). This is acceptable provided
that $R_{\rm eff}\approx1$ in units of lattice spacing. Here we
confirm that such a condition holds for the MoSe$_{2}$ line defects.
Nevertheless, we show that accounting for the effects of $R_{\rm eff}$
improves the agreement with the experiments beyond that reached by
\textcite{MoSe-17}. 

As in that reference, the SDS exponent $\alpha=(2-{\tilde{\xi}}_{c}^{2})^{2}/(8{\tilde{\xi}}_{c}^{2})$
is chosen to refer to the ${\tilde{\xi}}_{c}$ value for which there
is agreement between the specific $k$ intervals at which the $\gamma=s,c,c'$
branch-lines exponents ${\tilde{\zeta}}_{\gamma}(k)$ given in Eq. (\ref{equA3}) of Appendix \ref{APA}
are negative and the ARPES peaks distribution. For the $c$ and $s$ branch lines,
these intervals are $k \in [-2k_F + k_{Fc}^{\rm ex}, 2k_F - k_{Fc}^{\rm ex}]$ and
$k \in [-k_F + k_{Fs}^{\rm ex}, k_F - k_{Fs}^{\rm ex}]$, respectively. 
On the one hand, here $k_{Fc}^{\rm ex}$ is the experimental momentum that corresponds to the theoretical 
small momentum $k_{Fc}^0$ that controls the TLL and cross over to TLL regimes momentum width
considered in the discussions of Sec. \ref{SFPS}. Consistent with those discussions, $k_{Fc}^{\rm ex}/k_F$ is vanishing or very small. 
On the other hand, $k_{Fs}^{\rm ex}>k_{Fs}^0$ such that $k_{Fs}^{\rm ex}/k_F\approx 0.12$ rather refers to the 
experimental momenta $k = \pm (k_F - k_{Fs}^{\rm ex})$ at which the theoretical $s$ branch exponent vanishes.
Hence it is negative and positive for $k \in [-k_F + k_{Fs}^{\rm ex}, k_F - k_{Fs}^{\rm ex}]$ and 
$k \in [-k_F + k_{Fs}^{\rm ex},-k_F]$\, ;$[k_F - k_{Fs}^{\rm ex},k_F]$,
respectively. Indeed, only for negative exponent values does the theoretical $s$ branch line corresponds to ARPES peaks.
(See $s$ branch line exponent in Fig. \ref{figure5} for the value $\alpha =0.72$, for which, as discussed below, there 
is agreement between theory and experiments.) Finally, the $c'$ branch line exponent should be positive for its whole $k$ interval. 

The exponents in Eq. (\ref{equA3}) of Appendix \ref{APA} 
depend {\it both} on ${\tilde{\xi}}_c$ and momentum-dependent phase shifts
${\tilde{\Phi}}_{c,c}(\pm 2k_F,q)$ and ${\tilde{\Phi}}_{c,s}(\pm 2k_F,q')$. 
There is no apparent direct relation between the high-energy ARPES peaks 
distribution and the low-energy SDS. That the MQIM-HO contains the main microscopic mechanisms 
behind the 1D metallic states physics in the MoSe$_2$ line defects then requires 
that the $\alpha$ value that refers to the ${\tilde{\xi}}_{c}$ value for
which there is agreement with the high-energy ARPES peaks distribution
is also that measured within the low-energy angle integrated photoemission intensity.

We use in the expressions of the exponents ${\tilde{\zeta}}_{c}(k)$ and ${\tilde{\zeta}}_{c'}(k)$,
Eq. (\ref{equA3}) of Appendix \ref{APA}, the expression of the phase shift $2\pi{\tilde{\Phi}}_{c,c} (\pm 2k_F,q)$
in Eq. (\ref{PhaseShifts}) of that of Appendix, which includes the term ${\tilde{\Phi}}_{c,c}^{R_{\rm eff}} (k_r)$. We then find
that the parameters values that at electronic density
$n_{e}=2/3$ lead to agreement between the above intervals of the
$s$, $c$, and $c'$ branch lines {[}see Fig.~\ref{figure4}  (a){]} and the
line defects ARPES peaks distribution are $u=U/4t=0.18$, ${\tilde{\xi}}_c=0.655$,
$\alpha=0.72$, and $l=8$ for transfer integral $t=0.58$~eV. 

The corresponding $\gamma=c,c',s$ exponents ${\tilde{\zeta}}_{\gamma}(k)$
are plotted as a function of $k$ in Fig.~\ref{figure5} for different ${\tilde{\xi}}_{c}$
values and corresponding $\alpha=(2-{\tilde{\xi}}_{c}^{2})^{2}/(8{\tilde{\xi}}_{c}^{2})$
values. The ${\tilde{\xi}}_{c}$ value below which the effects of long-range interactions 
become stronger is ${\tilde{\xi}}_c^{\oslash}=1/\xi_c =0.731$.
The matching $\alpha=0.72$ value refers to ${\tilde{\xi}}_c=0.655$ and $R_{\rm eff}=1.01$
in units of lattice spacing and agrees with the estimated experimental uncertainty, $\alpha=0.75\pm0.05$
\citep{MoSe-17}. The prediction of \textcite{MoSe-17} that $\alpha=0.78$
lays in that uncertainty range, which confirms that the approximation of using 
$R_{\rm eff}\approx 0$ in the expression of the phase shift $2\pi{\tilde{\Phi}}_{c,c} (\pm 2k_F,q)$
is acceptable.

The room-temperature experimental SDS of the MoSe$_{2}$ line defects
is plotted in Fig. \ref{figure3} along with analytical lines for $\alpha=0.70,75,80$.
The theoretical SDS universal power-law behavior controlled by the exponent 
$\alpha$ in Fig. \ref{figure3} though only applies at very low energy, up to $\approx0.07$ eV. 
For larger energy values the SDS loses its universal power-law behavior, its
form becoming different and specific to each many-electron problem.

Comparison with the experimental
points for that energy range reveals that concerning the $\alpha=0.70,75,80$
theoretical lines the best agreement is reached at $\alpha=0.70$.
This is consistent with our correction from $\alpha=0.78$ to $\alpha=0.72$
improving the agreement. This is physically appealing, as one expects
that the effective range should not be smaller than one lattice spacing.

%%%%Conclusions

\section{Conclusions} \label{conclusions}

Confined states at TGBs in MoSe$_{2}$ were
shown to be well described by a three orbital TB model, which is robust
to the details of the defect hoppings. The presence of a single band
(per spin) at the Fermi level is consistent with experiments. 

Modeling the confined states as a 1D interacting electronic system unveils
a MQIM $(k,\omega)$-plane behavior with an effective range for the
charge fractionalized particle - charge hole mobile impurity interaction
that extends up to the lattice spacing, in excellent agreement with
ARPES measurements. 

The robustness and the properties found here for
1D confined states in MoSe$_{2}$ extend to the full semiconducting
TMD family, giving rise to a new paradigm where one-dimensionality
is protected by the two-dimensionality of the host material.

%%%%%%%%%%%%%%%%%%%%%%%%%%%%%%%%%%%%%%%%%%%%%%%%%%%%%%%%%%%%%%%%%%%%%%%%%
\begin{acknowledgments}
T.\v{C}. and J.M.P.C. thank Hai-Qing Lin for interesting discussions. E.C. is grateful to Pedro Ribeiro for valuable
insights regarding the absence of a bound state in the low energy
description. J.M.P.C. thanks Maria C. Asensio, Matthias Batzill, and Francisco Guinea
for illuminating discussions, Boston University's Condensed
Matter Theory Visitors Program for support, and the hospitality of MIT. 
We acknowledge the support from NSAF U1530401 and computational resources from CSRC (Beijing),
the Portuguese FCT through the Grant No.~UID/FIS/04650/2013, Grant
No.~UID/CTM/04540/2013, Grant No.~PTDC/FIS-MAC/29291/2017, and Grant
No.~SFRH/BSAB/142925/2018, and the NSFC Grant 11650110443.
\end{acknowledgments}

%%%%%%%%%%%%%%%%%%%%%%%%%%%%%%%%%%%%%%%%%%%%%%%%%%%%%%%%%%%%%%%%%%%%%%%%%

%%%%%%%%%%%%%%%%%%%%%%%%%%%%%%%%%%%%%%%%%%%%%%%%%%%%%%%%%%%%%%%%%%%%%%%%%%
\appendix

\section{Derivation of the continuum theory} \label{secap:continuum}

Consider the 3-band tight-binding Hamiltonian of TMDs\citep{xiao3bTB}
applied to the $y<0$ side of the TGB {[}see Fig.~\ref{fig:latticeTB}(a){]},
\begin{eqnarray}
\mathcal{H}=\sum_{\mathbf{k}}\hat{\psi}_{\mathbf{k}}^{\dagger}H(\mathbf{k})\hat{\psi}_{\mathbf{k}},\label{eq:H3b}
\end{eqnarray}
with $\hat{\psi}_{\bm{k}}^{\dagger}=(\hat{c}_{\bm{k},z^{2}}^{\dagger},\hat{c}_{\bm{k},xy}^{\dagger},\hat{c}_{\bm{k},x^{2}-y^{2}}^{\dagger})$,
and 
\begin{eqnarray}
H(\mathbf{k})=\left(\begin{array}{ccc}
h_{0} & h_{1} & h_{2}\\
h_{1}^{*} & h_{11} & h_{12}\\
h_{2}^{*} & h_{12}^{*} & h_{22}
\end{array}\right),\label{eq:Hk}
\end{eqnarray}
where 
\begin{eqnarray}
h_{0} & = & \epsilon_{1}+2t_{0}\cos{2\alpha}+4t_{0}\cos{\alpha}\cos{\beta}\nonumber \\
h_{11} & = & \epsilon_{2}+2t_{11}\cos{2\alpha}+(t_{11}+3t_{22})\cos{\alpha}\cos{\beta}\nonumber \\
h_{22} & = & \epsilon_{2}+2t_{22}\cos{2\alpha}+(t_{22}+3t_{11})\cos{\alpha}\cos{\beta}\nonumber \\
h_{1} & = & 2it_{1}\sin{2\alpha}+2it_{1}\sin{\alpha}\cos{\beta}-2\sqrt{3}t_{2}\sin{\alpha}\sin{\beta}\nonumber \\
h_{2} & = & 2t_{2}\cos{2\alpha}-2t_{2}\cos{\alpha}\cos{\beta}+2\sqrt{3}it_{1}\cos{\alpha}\sin{\beta}\nonumber \\
h_{12} & = & 2it_{12}\sin{2\alpha}-4it_{12}\sin{\alpha}\cos{\beta}\nonumber \\
 &  & +\sqrt{3}(t_{22}-t_{11})\sin{\alpha}\sin{\beta},\label{eq:helem}
\end{eqnarray}
$\alpha=k_{x}a/2$, and $\beta=\sqrt{3}k_{y}a/2$. The $K$ ($\tau=+1$)
and $K'$ ($\tau=-1$) points in the BZ are 
\begin{eqnarray}
\tau\mathbf{K}=(\tau\frac{4\pi}{3a},0),
\end{eqnarray}
where $\alpha$ and $\beta$ take the values $\alpha_{\tau}=\tau\frac{2\pi}{3}$,
$\beta_{\tau}=0$. The Taylor expansion to the second order around
$K$ and $K'$ points, reads: 
\begin{eqnarray}
 & H(\tau\mathbf{K}+\mathbf{q})=H(\tau\mathbf{K})+H_{\mathbf{q}}^{(1)}+H_{\mathbf{q}}^{(2)}+\mathcal{O}(aq)^{3}\nonumber \\
 & =H(\tau\mathbf{K})+q_{i}(\partial_{i}H)_{\tau\mathbf{K}}+\frac{1}{2}q_{i}q_{j}(\partial_{i}\partial_{j}H)_{\tau\mathbf{K}}+\mathcal{O}(aq)^{3}\nonumber \\
 & =\left(\begin{array}{ccc}
\eta_{0} & \eta_{1} & \eta_{2}\\
\eta_{1}^{*} & \eta_{11} & \eta_{12}\\
\eta_{2}^{*} & \eta_{12}^{*} & \eta_{22}
\end{array}\right)\,+a\left(\begin{array}{ccc}
u_{0} & u_{1} & u_{2}\\
u_{1}^{*} & u_{11} & u_{12}\\
u_{2}^{*} & u_{12}^{*} & u_{22}
\end{array}\right)\nonumber \\
 & +a^{2}\left(\begin{array}{ccc}
v_{0} & v_{1} & v_{2}\\
v_{1}^{*} & v_{11} & v_{12}\\
v_{2}^{*} & v_{12}^{*} & v_{22}
\end{array}\right)+\mathcal{O}(aq)^{3},\label{eq:expansion}
\end{eqnarray}
with 
\begin{eqnarray}
 & \eta_{0}=\epsilon_{1}-3t_{0},\hspace{1em}\eta_{11}=\epsilon_{2}-\frac{1}{2}(3t_{11}+3t_{22}),\nonumber \\
 & \eta_{22}=\epsilon_{2}-\frac{1}{2}(3t_{11}+3t_{22}),\nonumber \\
 & \eta_{1}=0,\hspace{1em}\eta_{2}=0,\hspace{1em}\eta_{12}=-i\tau3\sqrt{3}t_{12},\label{eq:eta}
\end{eqnarray}
\begin{eqnarray}
 & u_{0}=0,\hspace{1em}u_{11}=\frac{3\sqrt{3}}{4}\tau(t_{11}-t_{22})q_{x},\nonumber \\
 & u_{22}=\frac{3\sqrt{3}}{4}\tau(t_{22}-t_{11})q_{x},\nonumber \\
 & u_{1}=-\frac{3}{2}it_{1}q_{x}-\tau\frac{3\sqrt{3}}{2}t_{2}q_{y},\nonumber \\
 & u_{2}=\tau\frac{3\sqrt{3}}{2}t_{2}q_{x}-\frac{3}{2}it_{1}q_{y},\nonumber \\
 & u_{12}=\tau\frac{3\sqrt{3}}{4}(t_{22}-t_{11})q_{y},\label{eq:u}
\end{eqnarray}
and 
\begin{eqnarray}
 & v_{0}=\frac{3}{4}t_{0}q^{2},\hspace{1em}v_{11}=\frac{3}{16}[(3t_{11}+t_{22})q_{x}^{2}+(t_{11}+3t_{22})q_{y}^{2}],\nonumber \\
 & v_{22}=\frac{3}{16}[(t_{11}+3t_{22})q_{x}^{2}+(3t_{11}+t_{22})q_{y}^{2}],\nonumber \\
 & v_{1}=\frac{3}{4}t_{2}q_{x}q_{y}+i\tau\frac{3\sqrt{3}}{8}t_{1}(q_{x}^{2}-q_{y}^{2}),\nonumber \\
 & v_{2}=\frac{3}{8}t_{2}(q_{x}^{2}-q_{y}^{2})-i\tau\frac{3\sqrt{3}}{4}t_{1}q_{x}q_{y},\nonumber \\
 & v_{12}=\frac{3}{8}(t_{11}-t_{22})q_{x}q_{y}+i\tau\frac{3\sqrt{3}}{4}t_{12}q^{2}.\label{eq:v}
\end{eqnarray}

Diagonalizing the 0th order Hamiltonian in Eq.~\eqref{eq:expansion},
\begin{widetext}
\begin{eqnarray}
H(\tau\mathbf{K})=\left(\begin{array}{ccc}
\epsilon_{1}-3t_{0} & 0 & 0\\
0 & \epsilon_{2}-\frac{1}{2}(3t_{11}+3t_{22}) & -i\tau3\sqrt{3}t_{12}\\
0 & i\tau3\sqrt{3}t_{12} & \epsilon_{2}-\frac{1}{2}(3t_{11}+3t_{22})
\end{array}\right),
\end{eqnarray}
one obtains for the respective eigenvectors and eigenvalues,
\begin{eqnarray}
 &  & |\psi_{c}(\tau\mathbf{K})\rangle=|\tau\mathbf{K},d_{z^{2}}\rangle,\,~~~~~~~~~~~~~~~~~~~~~~~~~~~~~~~~~~~\epsilon_{c}=\epsilon_{1}-3t_{0}\nonumber \\
 &  & |\psi_{v}(\tau\mathbf{K})\rangle=\frac{1}{\sqrt{2}}\left[|\tau\mathbf{K},d_{x^{2}-y^{2}}\rangle+i\tau|\tau\mathbf{K},d_{xy}\rangle\right],~~~~\epsilon_{v}=\epsilon_{2}-\frac{1}{2}(3t_{11}+3t_{22})-3\sqrt{3}t_{12}\nonumber \\
 &  & |\psi_{h}(\tau\mathbf{K})\rangle=\frac{1}{\sqrt{2}}\left[|\tau\mathbf{K},d_{x^{2}-y^{2}}\rangle-i\tau|\tau\mathbf{K},d_{xy}\rangle\right],~~~~\epsilon_{h}=\epsilon_{2}-\frac{1}{2}(3t_{11}+3t_{22})+3\sqrt{3}t_{12}\,,\label{eq:3bLow1}
\end{eqnarray}
with the undersripts meaning: conduction band ($c$), valence band
($v$), and highest energy band ($h$).
\end{widetext}

The transformation matrix that diagonalizes $H(\tau\mathbf{K})$ reads
\begin{eqnarray}
U_{\tau}=\left(\begin{array}{ccc}
1 & 0 & 0\\
0 & -i\tau/\sqrt{2} & 1\tau/\sqrt{2}\\
0 & i\tau/\sqrt{2} & 1\tau/\sqrt{2}
\end{array}\right),
\end{eqnarray}
and the first-order matrix in the eigenbasis of $H(\tau\mathbf{K})$
is to be written as 
\begin{widetext}
\begin{eqnarray}
\Sigma^{(1)}(\mathbf{q})=U_{\tau}H^{(1)}(\mathbf{q})U_{\tau}^{-1} & = & \left(\begin{array}{ccc}
u_{0} & \frac{1}{\sqrt{2}}(u_{2}+i\tau u_{1}) & \frac{1}{\sqrt{2}}(u_{2}-i\tau u_{1})\\
\frac{1}{\sqrt{2}}(u_{2}+i\tau u_{1})^{*} & \frac{1}{2}(u_{22}+u_{11})+\tau\mathrm{Im}[u_{12}] & \frac{1}{2}(u_{22}-u_{11})-i\tau\mathrm{Re}[u_{12}]\\
\frac{1}{\sqrt{2}}(u_{2}-i\tau u_{1})^{*} & \frac{1}{2}(u_{22}-u_{11})+i\tau\mathrm{Re}[u_{12}] & \frac{1}{2}(u_{22}+u_{11})-\tau\mathrm{Im}[u_{12}]
\end{array}\right)\nonumber \\
 & = & a\left(\begin{array}{ccc}
0 & t_{vc}^{(1)}(\tau q_{x}-iq_{y}) & t_{ch}^{(1)}(\tau q_{x}+iq_{y})\\
t_{vc}^{(1)}(\tau q_{x}+iq_{y}) & 0 & t_{vh}^{(1)}(\tau q_{x}-iq_{y})\\
t_{ch}^{(1)}(\tau q_{x}-iq_{y}) & t_{vh}^{(1)}(\tau q_{x}+iq_{y}) & 0
\end{array}\right),
\end{eqnarray}
\end{widetext}

with 
\begin{eqnarray}
 & t_{vc}^{(1)}=\frac{3}{2\sqrt{2}}(\sqrt{3}t_{2}+t_{1}),~~~~t_{ch}^{(1)}=\frac{3}{2\sqrt{2}}(\sqrt{3}t_{2}-t_{1}),\nonumber \\
 & t_{vh}^{(1)}=\frac{3\sqrt{3}}{4}(t_{22}-t_{11}).
\end{eqnarray}
The second order correction to the Hamiltonian can be written as\begin{widetext}
\begin{align}
\Sigma^{(2)}(\mathbf{q})=U_{\tau}H^{(2)}(\mathbf{q})U_{\tau}^{-1}= & a^{2}\left(\begin{array}{ccc}
v_{0} & \frac{1}{\sqrt{2}}(v_{2}+i\tau v_{1}) & \frac{1}{\sqrt{2}}(v_{2}-i\tau v_{1})\\
\frac{1}{\sqrt{2}}(v_{2}+i\tau v_{1})^{*} & \frac{1}{2}(v_{22}+v_{11})+\tau\text{Im}v_{12} & \frac{1}{2}(v_{22}-v_{11})-i\tau\text{Re}v_{12}\\
\frac{1}{\sqrt{2}}(v_{2}-i\tau v_{1})^{*} & \frac{1}{2}(v_{22}-v_{11})+i\tau\text{Re}v_{12} & \frac{1}{2}(v_{22}+v_{11})-\tau\text{Im}v_{12}
\end{array}\right)\nonumber \\
= & a^{2}\left(\begin{array}{ccc}
\chi_{c}q^{2} & t_{vc}^{(2)}(q_{x}+i\tau q_{y})^{2} & t_{ch}^{(2)}(q_{x}-i\tau q_{y})^{2}\\
t_{vc}^{(2)}(q_{x}-i\tau q_{y})^{2} & \chi_{v}q^{2} & t_{vh}^{(2)}(q_{x}+i\tau q_{y})^{2}\\
t_{ch}^{(2)}(q_{x}+i\tau q_{y})^{2} & t_{vh}^{(2)}(q_{x}-i\tau q_{y})^{2} & \chi_{h}q^{2}
\end{array}\right),\label{2ndOrd}
\end{align}
\end{widetext}with 
\begin{align}
\chi_{c} & =\frac{3}{4}t_{0},~~~~\chi_{v}=\frac{3}{8}(t_{11}+t_{22}+\sqrt{3}t_{12}),\nonumber \\
\chi_{h} & =\frac{3}{8}(t_{11}+t_{22}-\sqrt{3}t_{12}),~~~~t_{vc}^{(2)}=\frac{3}{8\sqrt{2}}(t_{2}-\sqrt{3}t_{1})\nonumber \\
t_{ch}^{(2)} & =\frac{3}{8\sqrt{2}}(t_{2}-\sqrt{3}t_{1}),~~~~t_{vh}^{(2)}=\frac{3}{16}(t_{22}-t_{11})\,.\label{eq:2ndHopp}
\end{align}

The effective second order Hamiltonian of the lowest conduction and
highest valence bands is then given by 
\begin{eqnarray*}
 & H_{eff}(\mathbf{q})=P_{l}\left[H_{0}+\Sigma^{(1)}(\mathbf{q})+\Sigma^{(2)}(\mathbf{q})\right]P_{l}\\
 & +\sum_{l=c,v}\frac{|\psi_{l}\rangle\langle\psi_{l}|\Sigma^{(1)}(\mathbf{q})P_{h}\Sigma^{(1)}(\mathbf{q})|\psi_{l}\rangle\langle\psi_{l}|}{\epsilon_{l}-\epsilon_{h}}\\
 & +\sum_{\underset{m\neq n}{m,n=c,v}}\frac{|\psi_{m}\rangle\langle\psi_{m}|\Sigma^{(1)}(\mathbf{q})P_{h}\Sigma^{(1)}(\mathbf{q})|\psi_{n}\rangle\langle\psi_{n}|}{\epsilon_{F}-\epsilon_{h}}
\end{eqnarray*}
with $P_{l}=|\psi_{c}\rangle\langle\psi_{c}|+|\psi_{v}\rangle\langle\psi_{v}|$,
$P_{h}=|\psi_{h}\rangle\langle\psi_{h}|$, and $\epsilon_{F}=\frac{\epsilon_{c}+\epsilon_{v}}{2}$.
After straightforward manipulation, we obtain 
\begin{eqnarray}
H_{eff}^{(y<0)}(\mathbf{q}) & = & \left(\begin{array}{cc}
\epsilon_{c} & 0\\
0 & \epsilon_{v}
\end{array}\right)+at_{vc}^{(1)}\left(\begin{array}{cc}
0 & \tau q_{x}-iq_{y}\\
\tau q_{x}+iq_{y} & 0
\end{array}\right)\nonumber \\
 &  & +a^{2}\left(\begin{array}{cc}
\chi_{c}q^{2} & t_{vc}^{(2)}(q_{x}+i\tau q_{y})^{2}\\
t_{vc}^{(2)}(q_{x}-i\tau q_{y})^{2} & \chi_{v}q^{2}
\end{array}\right)\nonumber \\
 &  & +a^{2}\left(\begin{array}{cc}
\xi_{c}q^{2} & t_{vch}(q_{x}+i\tau q_{y})^{2}\\
t_{vch}(q_{x}-i\tau q_{y})^{2} & \xi_{v}q^{2}
\end{array}\right)\nonumber \\
 & = & v\hbar(q_{x}\tau_{3}\sigma_{x}+q_{y}\sigma_{y})+(\Delta+\delta\xi a^{2}q^{2})\sigma_{3}\nonumber \\
 &  & +\zeta a^{2}(q_{x}\tau_{3}\sigma_{x}-q_{y}\sigma_{y})\sigma_{x}(q_{x}\tau_{3}\sigma_{x}\nonumber \\
 &  & -q_{y}\sigma_{y})+(\epsilon_{F}+\xi q^{2})\sigma_{0},\label{eq:H2u}
\end{eqnarray}
where $\xi_{c}=\frac{t_{ch}^{2}}{\epsilon_{c}-\epsilon_{h}}$, $\xi_{v}=\frac{t_{vh}^{2}}{\epsilon_{v}-\epsilon_{h}}$,
$t_{vch}=\frac{t_{vh}t_{ch}}{\epsilon_{F}-\epsilon_{h}}$, $v=at_{vc}^{(1)}$,
$\Delta=\frac{\epsilon_{c}-\epsilon_{v}}{2}$, $\delta\xi=\frac{\xi_{c}-\xi_{v}}{2}+\frac{\chi_{c}-\chi_{v}}{2}$,
$\zeta=t_{vc}^{(2)}+t_{vch}$, $\xi=\frac{\chi_{c}+\chi_{v}+\xi_{c}+\xi_{v}}{2}$,
and $\tau\rightarrow\tau_{3}$. Apart from the constant and the electron-hole
asymmetry terms proportional to $\sigma_{0}$, there is also a trigonal
warping term proportional to $\zeta$, as well as the massive Dirac
Hamiltonian with a quadratic term. Estimates for MoSe$_{2}$ give\citep{xiao3bTB},
$v=5.6\times10^{5}\,\text{ms}{}^{-1}$, $2\Delta=1.44\,\text{eV}$,
$\delta\xi=-0.30\,\text{eV}$, $\zeta=9.4\,\text{meV}$, and $\xi=0.8\,\text{meV}$.

In order to obtain a low energy two-band model for the $y>0$ side
of the TGB {[}see Fig.~\ref{fig:latticeTB}(a){]}, we must recognize
that the two sides are related by a $y\rightarrow-y$ transformation.
This allows us to right the 3-band tight-binding Hamiltonian for $y>0$
exactly as in Eqs.~\eqref{eq:H3b}, \eqref{eq:Hk}, and~\eqref{eq:helem},
with the replacement $\beta\rightarrow-\beta$ in Eq.~\eqref{eq:helem}.
It should also be noted that the $y\rightarrow-y$ transformation
affects the atomic orbital basis ($d_{xy}\rightarrow-d_{xy}$), so
that the three component operator $\hat{\psi}_{\bm{k}}^{\dagger}$
in Eq.~\eqref{eq:H3b} is to be read on the $y>0$ side as $\hat{\psi}_{\bm{k}}^{\dagger}=(\hat{c}_{\bm{k},z^{2}}^{\dagger},\hat{c}_{\bm{k},-xy}^{\dagger},\hat{c}_{\bm{k},x^{2}-y^{2}}^{\dagger})$.

We want to compare the two sides of the TGB, so it is convenient to
use the same basis, which requires the transformation $(\hat{c}_{\bm{k},z^{2}}^{\dagger},\hat{c}_{\bm{k},-xy}^{\dagger},\hat{c}_{\bm{k},x^{2}-y^{2}}^{\dagger})\rightarrow(\hat{c}_{\bm{k},z^{2}}^{\dagger},\hat{c}_{\bm{k},xy}^{\dagger},\hat{c}_{\bm{k},x^{2}-y^{2}}^{\dagger})$
on the $y>0$ side. The unitary operator transforming between the
two basis is just $U=\text{diag}(1,-1,1)$, and the transformed Hamiltonian,
Taylor expanded to the second order near the $K$ ($\tau=+1$) and
$K'$ ($\tau=-1$) points, reads
\begin{eqnarray}
 & H(\tau\mathbf{K}+\mathbf{q})=H(\tau\mathbf{K})+H_{\mathbf{q}}^{(1)}+H_{\mathbf{q}}^{(2)}+\mathcal{O}(aq)^{2}\nonumber \\
 & =H(\tau\mathbf{K})+q_{i}(\partial_{i}H)_{\tau\mathbf{K}}+\frac{1}{2}q_{i}q_{j}(\partial_{i}\partial_{j}H)_{\tau\mathbf{K}}+\mathcal{O}(aq)^{3}\nonumber \\
 & =\left(\begin{array}{ccc}
\eta_{0} & -\eta_{1} & \eta_{2}\\
-\eta_{1}^{*} & \eta_{11} & -\eta_{12}\\
\eta_{2}^{*} & -\eta_{12}^{*} & \eta_{22}
\end{array}\right)+a\left(\begin{array}{ccc}
u_{0} & -u_{1} & u_{2}\\
-u_{1}^{*} & u_{11} & -u_{12}\\
u_{2}^{*} & -u_{12}^{*} & u_{22}
\end{array}\right)\nonumber \\
 & +a^{2}\left(\begin{array}{ccc}
v_{0} & -v_{1} & v_{2}\\
-v_{1}^{*} & v_{11} & -v_{12}\\
v_{2}^{*} & -v_{12}^{*} & v_{22}
\end{array}\right)+\mathcal{O}(aq)^{3},\label{eq:expansion2}
\end{eqnarray}
where the matrix elements $\eta$, $u$, and $v$, are the same as
in Eqs.~\eqref{eq:eta}, \eqref{eq:u}, and~\eqref{eq:v}, respectively,
with the replacement $q_{y}\rightarrow-q_{y}$ in Eqs.~\eqref{eq:u}
and~\eqref{eq:v}.

Diagonalizing the 0th order Hamiltonian in Eq.~\eqref{eq:expansion2},\begin{widetext}
\begin{eqnarray}
H(\tau\mathbf{K})=\left(\begin{array}{ccc}
\epsilon_{1}-3t_{0} & 0 & 0\\
0 & \epsilon_{2}-\frac{1}{2}(3t_{11}+3t_{22}) & +i\tau3\sqrt{3}t_{12}\\
0 & -i\tau3\sqrt{3}t_{12} & \epsilon_{2}-\frac{1}{2}(3t_{11}+3t_{22})
\end{array}\right),\label{eq:0thHk2}
\end{eqnarray}
one obtains
\begin{eqnarray}
 &  & |\psi_{c}(\tau\mathbf{K})\rangle=|\tau\mathbf{K},d_{z^{2}}\rangle,~~~~~~~~~~~~~~~~~~~~~~~\,~~~~~~~~~~~~\epsilon_{c}=\epsilon_{1}-3t_{0}\nonumber \\
 &  & |\psi_{h}(\tau\mathbf{K})\rangle=\frac{1}{\sqrt{2}}\left(|\tau\mathbf{K},d_{x^{2}-y^{2}}\rangle+i\tau|\tau\mathbf{K},d_{xy}\rangle\right),~~~~\epsilon_{h}=\epsilon_{2}-\frac{1}{2}(3t_{11}+3t_{22})+3\sqrt{3}t_{12}\nonumber \\
 &  & |\psi_{v}(\tau\mathbf{K})\rangle=\frac{1}{\sqrt{2}}\left(|\tau\mathbf{K},d_{x^{2}-y^{2}}\rangle-i\tau|\tau\mathbf{K},d_{xy}\rangle\right),~~~~\epsilon_{v}=\epsilon_{2}-\frac{1}{2}(3t_{11}+3t_{22})-3\sqrt{3}t_{12}.\nonumber \\
\label{eq:3bLow2}
\end{eqnarray}
\end{widetext}Comparing the atomic content of the two states $|\psi_{v}(\tau\mathbf{K})\rangle$
and $|\psi_{h}(\tau\mathbf{K})\rangle$ in Eq.~\eqref{eq:3bLow2}
with their counterparts in Eq.~\eqref{eq:3bLow1}, it is apparent
that a gap inversion occurs between the two as we cross the boundary.
This gap inversion is further discussed in the main text, Sec.~\ref{subsec:lowE3b}.

The effective Hamiltonian for $y>0$ in the subspace of the conduction
and valence bands may now be obtained in a similar way to the $y<0$
side. We first use the basis in Eq.~\eqref{eq:3bLow2} to write the
expanded Hamiltonian of Eq.~\eqref{eq:expansion2}, and then apply
exactly the same procedure as for the $y<0$ side after Eq.~\eqref{eq:3bLow1}.
We finally arrive at

\begin{eqnarray}
H_{eff}^{(y>0)}(\mathbf{q}) & = & v\hbar(q_{x}\tau_{3}\sigma_{x}-q_{y}\sigma_{y})+(\Delta+\delta\xi a^{2}q^{2})\sigma_{3}\nonumber \\
 &  & +\zeta a^{2}(q_{x}\tau_{3}\sigma_{x}+q_{y}\sigma_{y})\sigma_{x}(q_{x}\tau_{3}\sigma_{x}\nonumber \\
 &  & +q_{y}\sigma_{y})+(\epsilon_{F}+\xi q^{2})\sigma_{0},\label{eq:H2d}
\end{eqnarray}
which is exactly the same as Eq.~\eqref{eq:H2u} after the transformation
$q_{y}\rightarrow-q_{y}$. The parameters in Eq.~\eqref{eq:H2d}
are the same as in Eq.~\eqref{eq:H2u}.

\section{Some MQIM-HO useful expressions} \label{APA}

The spectra of the $\gamma =s,c,c'$ branch lines in the spectral-function expression, Eq. (\ref{Bkomega}), 
are given by,
\begin{eqnarray}
{\tilde{\omega}}_{s} (k) & = & {\tilde{\varepsilon}}_s (k) = \varepsilon_{s} (k) \leq 0\hspace{0.20cm}{\rm for}
\hspace{0.20cm} k = -q' \in [-k_F,k_F] 
\nonumber \\
{\tilde{\omega}}_c (k) & = & {\tilde{\varepsilon}}_c (\vert k\vert + k_F)\leq 0
\hspace{0.20cm}{\rm for}
\nonumber \\
k & = & k_c = -{\rm sgn}\{k\} k_F - q \in [-k_F,k_F] 
\nonumber \\
{\tilde{\omega}}_{c'} (k) & = & {\tilde{\varepsilon}}_c (\vert k\vert - k_F) \leq 0
\hspace{0.20cm}{\rm for}
\nonumber \\
k & = & k_{c'} = {\rm sgn}\{k\} k_F - q \in [-3k_F,3k_F] \, ,  
\label{equA1}
\end{eqnarray}
where ${\tilde{\varepsilon}}_s (q')$ and ${\tilde{\varepsilon}}_c (q)$ are the $s$ and
$c$ particle energy dispersions, respectively, given below in Eq. (\ref{equA2}). The
spectra, Eq. (\ref{equA1}), are plotted within the MQIM-HO in 
Fig. \ref{figure4} (a) as a function of the excitation momentum $k$ for 
$u=U/4t=0.18$, transfer integral $t=0.58$ eV, and electronic density $n_{e}=2/3$.

As discussed in Sec. \ref{SFPS}, the charge particle - charge hole mobility impurity interaction gives rise to a
slight renormalization of the $c$ band energy dispersion. Within the MQIM-HO it is estimated to lead to,
\begin{eqnarray}
{\tilde{\varepsilon}}_c (q) & = & \left(1 + \beta_c\,\theta_c\right)\varepsilon_c (q)\hspace{0.20cm}{\rm for}\hspace{0.20cm} 
q \in ]-2k_F,2k_F[  
\nonumber \\
{\tilde{\varepsilon}}_s (q') & = & \varepsilon_s (q') \hspace{0.20cm}{\rm for}\hspace{0.20cm} 
q' \in ]-k_F,k_F[  \, ,
\label{equA2}
\end{eqnarray}
where $\beta_c = {1\over\xi_c}\left(1 - {\xi_c \over\sqrt{2}}\right)$ and  
$\theta_c = 1$ for ${\tilde{\xi}}_c\in ]1/2,1[$ and $\theta_c =\left({\xi_c - {\tilde{\xi}}_c\over \xi_c - 1}\right)$ 
for ${\tilde{\xi}}_c\in ]1,\xi_c[$. Here the $s$ band energy dispersion,
which remains invariant under the universal transformation, was also given. 
The 1DHM dispersions $\varepsilon_c (q)$ and $\varepsilon_s (q')$ in Eq. (\ref{equA1}) are defined by \textcite{MoSe-17}.

The $\gamma = c,c',s$ exponents ${\tilde{\zeta}}_{\gamma} (k)$ in the spectral function, Eq. (\ref{Bkomega}), 
plotted in Fig. \ref{figure5} for $u=0.18$, $n_{e}=2/3$, and $l=8$ read,
\begin{eqnarray}
{\tilde{\zeta}}_c (k) & = & -{1\over 2} + \sum_{\iota=\pm1}\left({{\tilde{\xi}}_c\over 4} - {\tilde{\Phi}}_{c,c}(\iota 2k_F,q)\right)^2  
\hspace{0.20cm}{\rm where}
\nonumber \\
k & = & \in [-k_F+k_{Fc}^0,k_F-k_{Fc}^0] 
\nonumber \\
q & = & -{\rm sgn}\{k\} k_F - k \in [-2k_F+k_{Fc}^0,-k_F] \hspace{0.20cm}{\rm and}
\nonumber \\
& = & -{\rm sgn}\{k\} k_F - k \in [k_F,2k_F-k_{Fc}^0] 
\nonumber \\
{\tilde{\zeta}}_{c'} (k) & = & -{1\over 2} + \sum_{\iota=\pm1}\left({{\tilde{\xi}}_c\over 4} - {\tilde{\Phi}}_{c,c}(\iota 2k_F,q)\right)^2  
\hspace{0.20cm}{\rm where}
\nonumber \\
k & = & \in [-3k_F+k_{Fc}^0,3k_F-k_{Fc}^0] 
\nonumber \\
q & = & {\rm sgn}\{k\} k_F - k \in [-2k_F+k_{Fc}^0,k_F] \hspace{0.20cm}{\rm and} 
\nonumber \\
& = & {\rm sgn}\{k\} k_F - k \in [-k_F,2k_F-k_{Fc}^0] 
\nonumber \\
{\tilde{\zeta}}_{s} (k) & = & -1 + \sum_{\iota=\pm1}\left(- {\iota\over 2{\tilde{\xi}}_c} - {\tilde{\Phi}}_{c,s}(\iota 2k_F,q')\right)^2  
\hspace{0.20cm}{\rm where}
\nonumber \\
k & \in & [-k_F+k_{Fs}^0,k_F-k_{Fs}^0] 
\nonumber \\
q' & = & -k \in  [-k_F+k_{Fs}^0,k_F-k_{Fs}^0] \, . 
\label{equA3}
\end{eqnarray}

The renormalization of the phase shifts $2\pi{\tilde{\Phi}}_{c,s}(\pm2k_{F},q')$ and $2\pi{\tilde{\Phi}}_{c,c}(\pm 2k_{F},q)$ 
appearing in the exponents expressions, Eq. (\ref{equA3}),
under the $\xi_c\rightarrow {\tilde{\xi}}_c$ transformation of \textcite{MoSe-17} leads to,
\begin{eqnarray}
2\pi{\tilde{\Phi}}_{c,s} (\pm 2k_F,q') & = & {{\tilde{\xi}}_c\over\xi_c}\,2\pi\Phi_{c,s} (\pm 2k_F,q') 
\nonumber \\
2\pi{\tilde{\Phi}}_{c,c} (\pm 2k_F,q) & = & 2\pi{\tilde{\Phi}}_{c,c}^{{\tilde{a}}} (\pm 2k_F,q) + 2\pi{\tilde{\Phi}}_{c,c}^{R_{\rm eff}} (k_r)
\nonumber \\
2\pi{\tilde{\Phi}}_{c,c}^{{\tilde{a}}} (\pm 2k_F,q) & = & 
{\xi_c\over {\tilde{\xi}}_c}{({\tilde{\xi}}_c -1)^2\over (\xi_c -1)^2}\,2\pi\Phi_{c,c} (\pm 2k_F,q) \, ,
\label{PhaseShifts}
\end{eqnarray}
for $q' \in [-k_F,k_F]$, $q \in [-2k_F^{+},2k_F^{-}]$, and 
$\vert k_r\vert = \vert q\mp 2k_F\vert \in [0,4k_F[$. Here $k_r^0=2\pi/L$ and
$k_r = (q\mp 2k_F)$ is the relative momentum of the charge particle at the $c$ band Fermi points $\pm 2k_F$ 
and charge hole mobile impurity of $c$ band momentum $q \in [-2k_F^{+},2k_F^{-}]$ and $\Phi_{c,s} (\pm 2k_F,q')$ and
$\Phi_{c,c} (\pm 2k_F,q)$ are 1DHM phase shifts. 

The spin-particle phase shifts remain invariant under the MQIR-LR transformation and are given by,
\begin{eqnarray}
{\tilde{\Phi}}_{s,s} (\iota k_F,q') & = & {\iota (\xi_s -1)(\xi_s + (-1)^{\delta_{q,\iota k_F}})\over 2\xi_s} 
\nonumber \\
{\rm for} & & q' \in  [-k_F,k_F]
\nonumber \\
{\tilde{\Phi}}_{s,c}(\iota k_F,q) & = & -{\iota\xi_s\over 4} \hspace{0.20cm}
{\rm for}\hspace{0.20cm}q \in  [-2k_F,2k_F] \, ,
\label{equA4}
\end{eqnarray}
where $\xi_s = \sqrt{2}$ and $\iota=\pm 1$.\\ \\

\bibliographystyle{apsrev4-1}
\bibliography{biblio_MoSe2-TGB}

\end{document}